\documentclass[twocolumn]{aastex6}

\usepackage{amsmath}

\def \halpha{H$\alpha$}
\def \Msol{{M}_{\odot}}

\def \logm{\log(M/\Msol)}

\def \h2{{\rm H_{2}}}
\def \oabund{12+\log({\rm O/H})}

\def \halpha{H$\alpha$}

\def \niiha{[N{\scriptsize ~II}]/H$\alpha$}

\def \niiniihatot{[N{\scriptsize ~II}]/([N{\scriptsize ~II}]+H$\alpha$)$_{\rm tot}$}
\def \niipha{([N{\scriptsize ~II}]+H$\alpha$)}
\def \niiphatot{([N{\scriptsize ~II}]+H$\alpha$)$_{\rm tot}$}
\def \oiiihb{[O{\scriptsize ~III}]/H$\beta$}
\def \loiiihb{$\log($[O{\scriptsize ~III}]/H$\beta$)}

\def \oii{[O{\scriptsize ~II}]}
\def \oiii{[O{\scriptsize ~III}]}

\def \nii{[N{\tiny\,II}]}

\def \logL{\log(L)}

\def \dn4000{D_{{\rm n}}(4000) }
\newcommand{\ergPerSecondPerCM}{{\rm erg}\,{\rm s}^{-1}\,{\rm cm}^{-2}}

\begin{document}


\title{Empirical modeling of the Redshift evolution of the \nii/\halpha~ratio \\for galaxy redshift surveys}


\author{
Andreas L. Faisst$^{1}$,
Daniel Masters$^{1}$,
Yun Wang$^{1,2}$,
Alexander Merson$^{1,3}$,
Peter Capak$^{1}$,
Sangeeta Malhotra$^{4}$, and
James E. Rhoads$^{4}$
}





\affil{
$^1$IPAC, Mail Code 314-6, California Institute of Technology, 1200 East California Boulevard, Pasadena, CA 91125, USA\\
$^2$Homer L. Dodge Department of Physics \& Astronomy, University of Oklahoma, 440 W Brooks Street, Norman, OK 73019, USA\\
$^3$Jet Propulsion Laboratory, California Institute of Technology, 4800 Oak Grove Drive, Pasadena, CA 91109, USA\\
$^4$School of Earth and Space Exploration, Arizona State University, Tempe, AZ 85287, USA}



\email{anfaisst@gmail.com}

\begin{abstract}

	We present an empirical parameterization of the \niiha\ flux ratio as a function of stellar mass and redshift valid at $0 < z < 2.7$ and $8.5 < \logm < 11.0$. This description can easily be applied to \textit{(i)} simulations for modeling \nii$\lambda6584$ line emission, \textit{(ii)} deblend \nii\ and \halpha\  in current low-resolution grism and narrow-band observations to derive intrinsic \halpha\ fluxes, and \textit{(iii)} to reliably forecast the number counts of \halpha\ emission-line galaxies for future surveys, such as those planned for \emph{Euclid} and the \emph{Wide Field Infrared Survey Telescope} (\emph{WFIRST}).
	Our model combines the evolution of the locus on the Baldwin, Phillips \& Terlevich (BPT) diagram measured in spectroscopic data out to $z\sim2.5$ with the strong dependence of \niiha\ on stellar mass and \oiiihb\ observed in local galaxy samples.
    We find large variations in the \niiha\ flux ratio at a fixed redshift due to its dependency on stellar mass; hence, the assumption of a constant \nii\ flux contamination fraction can lead to a significant under- or overestimate of \halpha\ luminosities. Specifically, measurements of the intrinsic \halpha\ luminosity function derived from current low-resolution grism spectroscopy assuming a constant $29\%$ contamination of \nii\ can be overestimated by factors of $\sim8$ at $\logL>43.0$ for galaxies at redshifts $z \sim 1.5$. This has implications for the prediction of \halpha\ emitters for \emph{Euclid} and \emph{WFIRST}.
    We also study the impact of blended \halpha\ and \nii\ on the accuracy of measured spectroscopic redshifts.

\end{abstract}




\section{Introduction}
		
        The \halpha\ nebular emission line at rest-frame $6563\,{\rm \AA}$ is the most important feature that will be detected by the near-infrared grisms of the upcoming space-based missions \emph{Euclid} \citep{LAUREIJS11,VAVREK16} and \emph{WFIRST} \citep[][]{DRESSLER12,GREEN12b,SPERGEL15}. By measuring redshifts to tens of millions of \halpha\ emitters at $1\lesssim z \lesssim 2$ with low-resolution grism spectroscopy, these surveys will use baryon acoustic oscillations \citep[BAO, e.g.,][]{BLAKE03,SEO03} and redshift space distortions \citep[RSD,][]{KAISER87} analyses to constrain the expansion history of the universe and the growth of structure. Together, these probes will put strong constraints on the nature of dark energy \citep{GUZZO08,WANG08}. 
        
The \halpha~emission of a galaxy depends principally on its star formation rate (SFR), while the \halpha~equivalent-width (EW) is proportional to the specific SFR (sSFR\,=\,SFR/$M$, a proxy for the rate of stellar mass increase). The \halpha\ emission is, therefore, not only an important tool to study cosmology but also a direct probe of the statistics of cosmic star formation via the \halpha\ luminosity function (LF). The LF, in turn, informs predictions of the number counts of galaxies that will be found in the grism surveys of \emph{Euclid} and \emph{WFIRST}.
   
Current measurements of \halpha\ across large redshift ranges up to $z\sim2$ and over large area on sky come from blind searches in low-resolution \emph{Hubble Space Telescope} (\emph{HST}) grism observations \citep[][]{ATEK10,BRAMMER12,VANDOKKUM13}. 
Specifically, the best current constraints on the \halpha\ luminosity function over the redshifts of interest for \emph{Euclid} and \emph{WFIRST} ($0.4 < z < 2.5$) come from the HST WFC3 Infrared Spectroscopic Parallel Survey \citep[WISPS,][]{ATEK10}. However, these WFC3 grism spectra do not resolve \halpha\ from the neighboring \nii\ lines at rest-frame $6548\,{\rm \AA}$ and $6584\,{\rm \AA}$.
The resulting uncertainty in the \nii/\halpha\ ratio (in the following, \niiha$\,\equiv\,$\nii$\lambda6584/$\halpha)\footnote{We assume \nii$_{6548} = \frac{1}{3}$\nii$\lambda6584$ \citep{ACKER89}, and therefore \nii$\lambda\lambda6548,6584/$\halpha$\,=\frac{4}{3}$\niiha.} for the sources measured by WFC3 translates into significant uncertainty in the derived intrinsic \halpha\ LF and therefore has a direct impact on shaping future large surveys. Commonly, a constant \nii\ flux contamination fraction of $29\%$ is assumed for such grism surveys according to the average value measured in the local universe for galaxies with \halpha\ EW less than $200\,{\rm \AA}$ \citep{COLBERT13,MEHTA15}. However, such an assumption can introduce luminosity dependent biases because of the dependence of the \niiha\ ratio on several galaxy parameters including the redshift and stellar mass \citep[see also discussion in][]{POZZETTI16}.

    The dark energy figure-of-merit of both \emph{Euclid} and \emph{WFIRST} is very sensitive to the number density of \halpha\ emitting galaxies. Due to the sharp exponential fall-off at the bright end of the \halpha\ LF that will be probed by these surveys, an uncertainty of a factor of 3 in the \niiha\ flux ratio, for example due to blending, could translate into an uncertainty of a factor of up to $10$ in the number counts of bright \halpha\ emitting galaxies in the worst case.
    Furthermore, similar to the grism surveys, most of the \halpha\ lines detected by \emph{Euclid} and some detected by \emph{WFIRST} will be blended with \nii, degrading the accuracy of its measured \halpha\ LF. In addition, incorrect (or no) de-blending of these two lines can result in a systematic offset of the \halpha\ line centroid of up to $300\,{\rm km\,s}^{-1}$ (depending on the actual \niiha\ ratio) and therefore directly affect the accuracy of redshift measurements (Section~\ref{sec:centroid}).    
    
\begin{figure}
\centering
\includegraphics[width=1.0\columnwidth, angle=0]{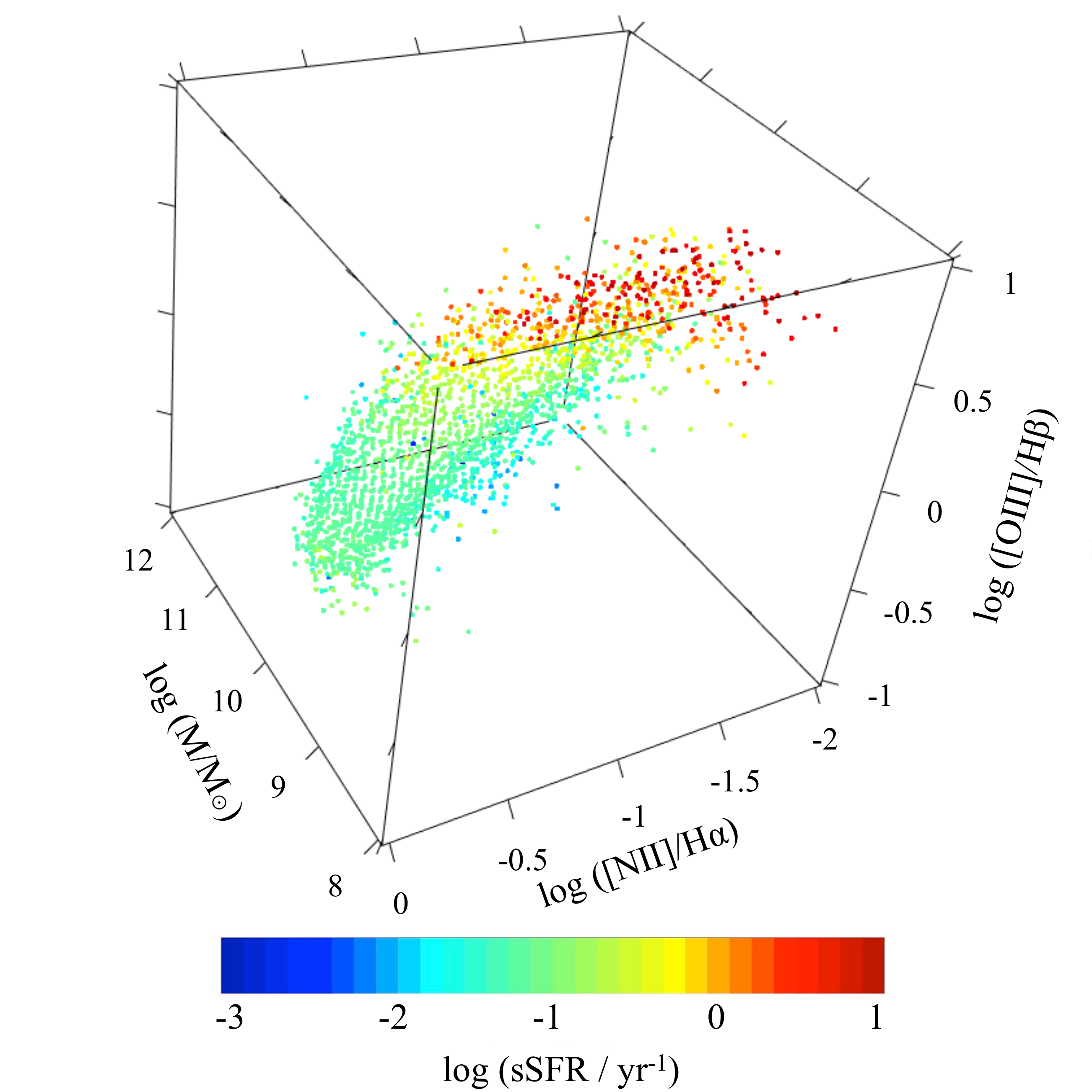}
\caption{Dependences between \niiha, \oiiihb, stellar mass, and sSFR (indicated in color) based on the sample of local galaxies in SDSS. The data is median-binned for visual purposes. An interactive three-dimensional version of this plot built with \texttt{plotly} is available in the online Journal (\texttt{plotly}: \url{https://plot.ly/}; a version is also available on the author's website: \url{http://www.astro.caltech.edu/afaisst/3dplot/plotly_3dplot1.html}). The data used to create this figure are available online.
\label{fig:bpt}}
\end{figure}

   A significant variation (by factors of $3\!-\!5$) in the \niiha~ratio is expected from the large parameter space spanned by galaxies on the ``Baldwin, Phillips \& Terlevich'' \citep[BPT,][]{BALDWIN81} diagram, which has \niiha\ as the abscissa. This variation is known to be linked to galaxy properties such as metallicity, SFR, and nitrogren-to-oxygen (N/O) ratio, and, due to changing galaxy demographics, the population averaged \niiha~ratio is expected to vary as a function of redshift \citep[e.g.,][]{KEWLEY13,MASTERS14,STEIDEL14,SHAPLEY15,MASTERS16,STROM17,KASHINO17}.
    Simulations may be used to predict the \niiha~ratio as a function of galaxy parameters \citep[e.g.,][]{HIRSCHMANN17}; however, large uncertainties can arise due to different ingredients and assumptions. Other studies use the  \halpha\ EW in local galaxy samples as prior for the \niiha\ ratio \citep[see e.g.,][]{VILLAR08,SOBRAL09,LY11,LEEJANICE12,SOBRAL13}.
       
     Here we outline an approach to constrain the expected \niiha\ ratios for upcoming surveys using empirical trends in the BPT diagram. Our approach is motivated by the correlations in this diagram illustrated by \citet{MASTERS16} and \citet{FAISST16c} \citep[see also][]{BRINCHMANN08}. \citet{MASTERS16} showed the strong correlation of both stellar mass and SFR density with position on the BPT diagram. Galaxies form a tight locus in the BPT diagram at $z\sim0$; however, the position of this locus is known to evolve with redshift \citep{ERB06,MASTERS14,STEIDEL14,SHAPLEY15}, which is driven by the changing average sSFR with cosmic time \citep{DADDI07a,ELBAZ07,NOESKE07,LILLY13}, metallicity \citep{LY16b}, ionization parameter \cite[e.g.,][]{NAKAJIMA14}, and electron density of the galaxies.
          
     Here we illustrate that stellar mass and redshift together put strong constraints on the \niiha\ ratio. We parameterize the evolution in the \niiha-mass relation with redshift, which allows us to accurately predict the \niiha\ ratio for galaxies with known stellar mass and redshift.
     The model we present can be used to
     \begin{itemize}
     \item Accurately deblend \nii\ and \halpha\ in low-resolution spectroscopic surveys and narrow-band photometric observations,
     \item Improve the fidelity of the forecasts for the number counts of \halpha\ emitters that will be detected by \textit{Euclid} and \emph{WFIRST}.
     \end{itemize}
     
     First, we outline the idea and motivation of our approach in Section~\ref{sec:idea}.
     In Section~\ref{sec:model}, we describe our empirical model that parameterizes the \niiha\ ratio as a function of stellar mass and redshift. The observational data that is feeding our model is presented in Section~\ref{sec:data}. In the following sections we derive our model and present the final parameterization in Section~\ref{sec:final}. 
     In Section~\ref{sec:implications}, we study in detail the implications of our model on \textit{(i)} the data interpretation of current surveys (Section~\ref{sec:datainterp}), \textit{(ii)} the \nii\ contamination for \emph{Euclid} and \emph{WFIRST} (Section~\ref{sec:n2contamination}), \textit{(iii)} redshift measurement from blended \halpha\ and \nii\ lines (Section~\ref{sec:centroid}), and \textit{(iv)} the number count predictions for future surveys based on current \halpha\ LF determinations (Section~\ref{sec:half}).
 
    Throughout this paper we assume a flat cosmology with $\Omega_{\Lambda,0} = 0.7$, $\Omega_{m,0} = 0.3$, and $h = 0.7$. Furthermore, all stellar masses and star-formation rates (SFR) are scaled to a \citet[][]{CHABRIER03} initial mass function (IMF) and all magnitudes are quoted in AB \citep{OKE74}.

\begin{figure}
\centering
\includegraphics[width=1.0\columnwidth, angle=0]{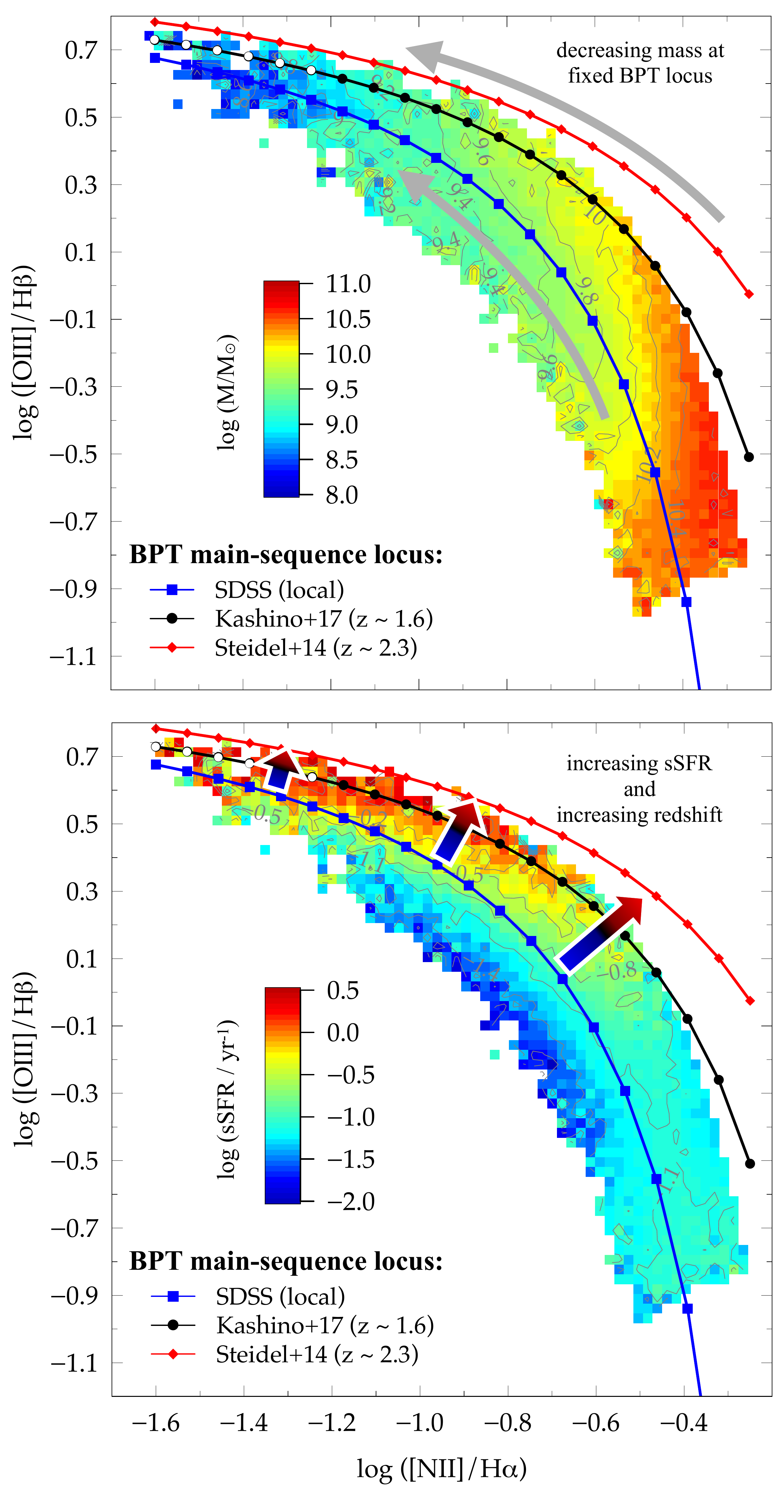}
\caption{Slices through Figure~\ref{fig:bpt} showing the dependence in stellar mass (\textit{top}) and sSFR (\textit{bottom}) in the local BPT diagram (fit indicated by blue line). High-redshift galaxies occupy sub-regions of the local BPT diagram at higher sSFR (perpendicular to the local BPT locus) as shown by the measured ``BPT main-sequence'' loci at $z\sim1.6$ \citep[extrapolated to $\log($\niiha$)=-1.6$,][]{KASHINO17} and $z\sim2.3$ \citep{STEIDEL14,STROM17}. Stellar mass runs nearly perpendicular to the loci indicating a strong mass dependence. This allows a unique description of emission line ratios as a function of redshift and stellar mass, which is the cornerstone of our model described in the text.
\label{fig:bpt2}}
\end{figure}

\section{Background and Motivation} \label{sec:idea}

	We model the \niiha~ratio as a function of redshift and stellar mass, two key observational quantities.
	While other methods to constrain \nii\ contamination using optical emission lines could potentially be more accurate \citep[e.g., involving \oiii\ and \oii, ][]{DELOSREYES15}, these cannot be applied to, e.g., narrow-band observations (lacking spectroscopic follow-up) as well as the large samples of future large-area surveys such as with \emph{WFIRST} or \emph{Euclid} due to their relatively low spectroscopic line sensitivity. However, at \emph{WFIRST} and \emph{Euclid} depths\footnote{\emph{Euclid} will obtain $Y-$, $J-$, and $H-$band imaging down to $24\,{\rm AB}$ for a $5\sigma$ point source and \emph{WFIRST} will reach $\sim26.5\,{\rm AB}$ \citep[e.g.,][]{GEHRELS15}.}, stellar masses will be accurately determined down to at least $\logm = 9.0-9.5$ by covering the $4000\,{\rm \AA}$ break up to $z=2$ with $Y-$, $J-$, and $H-$band imaging \citep[][]{LAUREIJS11,GEHRELS15}. This justifies our approach of using redshift and stellar mass as main quantity to derive the \niiha\ ratio.
	Moreover, stellar mass and redshift are well-constrained in the semi-analytical models that are often used to estimate population statistics for \halpha\ emitters \citep[e.g.,][]{ORSI10, MERSON18}. 
	The evolution of the \niiha\ ratio is empirically constrained both by trends seen in the local SDSS sample as well as by measured \niiha-stellar mass relations out to $z\sim2$ in the literature. Since the \niiha\ ratio is a gas-phase metallicity indicator \citep[e.g.,][]{PETTINI04}, the \niiha$-$mass relation is effectively the galaxy mass$-$metallicity (MZ) relation, which is known to evolve with redshift \citep[e.g.,][]{SAVAGLIO05,ERB06,MAIOLINO08,LILLY13,MAIER15,SALIM15}. 

	Physical galaxy properties such as stellar mass, sSFR, and relative abundance ratios are strongly correlated with nebular emission line ratios. This is shown in Figure~\ref{fig:bpt}, a 3-dimensional version of the BPT-diagram connecting the line ratios \niiha\ and \oiiihb\ with the more easily accessible observables stellar mass ($M$) and sSFR (color coded).
    Figure~\ref{fig:bpt2} shows projections of the 3-dimensional figure to better visualize the dependencies with stellar mass (top) and sSFR (bottom). The fitted locus of local galaxies \citep{KEWLEY08} is indicated with a blue line.
    
    Trends in the local SDSS data with SFR and stellar mass reflect changes seen in the galaxy population at high redshift.
    The BPT locus \emph{systematically} shifts with redshift, possibly connected to the overall increase in the global SFR of galaxies.
    The measured BPT loci of galaxies at $z\sim1.6$ \citep{KASHINO17} and $z\sim2.3$ \citep{STEIDEL14} are indicated in the upper and lower panels of Figure~\ref{fig:bpt2}, illustrating the pronounced shift in the BPT locus.
    Indeed, the position on the BPT diagram is very effective to select ``high-redshift analogs'', which are a rare sub-sample of local galaxies that resemble high-redshift galaxies in photometric and spectroscopic properties \citep[such as high sSFR or \halpha\ equivalent-width, e.g.,][]{CARDAMONE09,HU09,STANWAY14,LY15,FAISST16c,GREIS16,ERB16,Ly16}.
   
    The locus on the BPT diagram of (roughly) constant sSFR is similar to the locus of the main-sequence on the stellar mass vs. SFR plane. At different redshifts, galaxies populate different distributions of sSFR \citep[e.g., compilations by][]{LILLY13,SPEAGLE14} and stellar mass ranges, which let the slope and normalization of the main-sequence change across cosmic time. Similarly, galaxies on a ``BPT main-sequence'' run through a range in \niiha, and \oiiihb, and stellar masses (see upper panel of Figure~\ref{fig:bpt2}) at an sSFR distribution most likely for their redshift. \citet{MASTERS16} identified stellar mass and its link to the nitrogen-to-oxygen (N/O) abundance ratio as the main driver for these dependencies, as these quantities strongly vary approximately perpendicular to lines of constant sSFR, i.e., the BPT main-sequence. Once the BPT main-sequence is identified, the \niiha~(and \oiiihb) can therefore be uniquely determined from the stellar mass of a galaxy.
    
    The steps for creating our model are therefore as follows:
\begin{enumerate}
\item Parameterize the BPT main-sequence as a function of redshift as $O3(N2,z)$,
\item Parameterize the mass dependence on the BPT diagram, i.e., $M(O3,N2)$,
\item Parameterize the \niiha\ ratio as a function of stellar mass and redshift, i.e., $N2(M, z)$, by reversing $M(N2,z)$.
\end{enumerate}    
     Here, and in the following, we adopt the definitions $N2=\log($\niiha) and $O3~=~\log($\oiiihb).
    Note, that a parameterization of $N2(M,z)$ can also be derived by directly reversing the observed relation between stellar mass and \niiha\ (which is proportional to the stellar mass $-$ gas-phase metallicity relation). It leads indeed to similar results, however, with a larger uncertainty and ambiguity caused by the \oiiihb\ dependence of the \niiha\ vs. stellar mass relation (see Figure~\ref{fig:bpt}). With our approach of parameterizing the entire mass dependence of the BPT diagram, we take this secondary dependence into account, which results in a more accurate and comprehensive model. Moreover this would also allow us to predict the \oiiihb\ ratios in addition to \niiha\ for a given stellar mass and redshift.

\section{Empirical model for the evolution of line-ratios}\label{sec:model} 

\begin{figure*}
\centering
\includegraphics[width=2.1\columnwidth, angle=0]{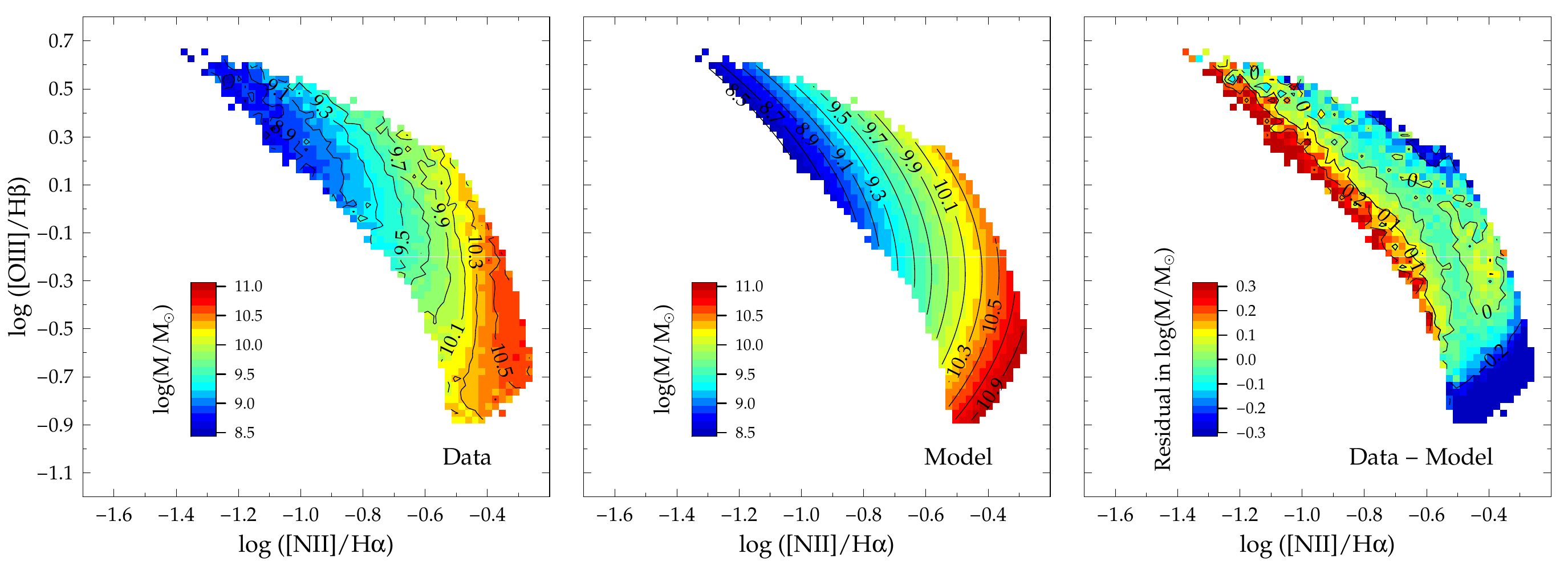}
\caption{Results of the multi-dimensional fit of $M(O3,N2)$ (Equation~\ref{eq:mass}) to the local galaxy sample from SDSS (\textit{left:} data; \textit{middle:} best-fit model; \textit{right:} residual). For the most parts, our model is able to recover stellar masses to better than $0.1\,{\rm dex}$. The best fit parameters are given in Section~\ref{sec:mass}.
\label{fig:bptfit}}
\end{figure*}

	\subsection{Local and high-redshift data}\label{sec:data}
    Our model is based on the data of $191\,409$ local galaxies selected from the Sloan Digital Sky Survey \citep[SDSS,][]{YORK00} using the web-based DR 12 \citep{ALAM15} query tool\footnote{\url{http://skyserver.sdss.org/dr12/en/tools/search/sql.aspx}}, combined with the observed BPT locus evolution at $0 < z < 2.5$ \citep{STEIDEL14,MASTERS14,SHAPLEY15,KASHINO17}.
    
    The stellar masses and emission line measurements in the SDSS catalog are taken from the \emph{Galspec} products provided in the MPA-JHU value added catalog based on the methods of \citet{KAUFFMANN03}, \citet{BRINCHMANN04}, and \citet{TREMONTI04}. Specifically, the stellar masses are derived from fits to the SDSS \emph{ugriz} total galaxy photometry assuming an exponentially declining star formation history and bursts. In addition, the photometry is corrected for the small contribution of nebular emission using the spectra. The model grids for the fitting are described in \citet{KAUFFMANN03} and a \citet{KROUPA01} IMF is assumed, which has been converted to a \citet{CHABRIER03} IMF.
    The local galaxies are selected to have a signal-to-noise ratio (S/N) $>5$ in the \halpha\ emission line. We note that different S/N thresholds do not have an impact on the subsequent analysis and result. Also, we do not impose a S/N limit on other optical emission lines in order to prevent our sample from any selection bias \citep[e.g.,][]{SALIM14}. Galaxies with a significant contribution of an active galactic nucleus (AGN) as suggested by the line ratios on the BPT diagram are removed \citep[about 15\%,][]{KEWLEY01,KAUFFMANN03,BRINCHMANN04}. The latter introduces an artificial upper boundary on galaxies on the BPT diagram at high \niiha\ and \oiiihb\ values, but the underlying trends in the distribution are not affected. In addition, we restrict our SDSS sample to $z>0.05$ in order to minimize the effect of the finite fiber aperture ($3\arcsec$ for the SDSS spectra), such that the $3\arcsec$ fiber covers at least the central $\sim 1.5\,{\rm kpc}$ of the galaxy. The SQL commands for this selection are provided in Appendix~\ref{app:sql}.
    
    In order to test the redshift dependence of our model, we make use of measurements of \niiha, \oiiihb, and stellar mass at higher redshifts as presented in the literature. These include galaxies at $z\sim1.6$ \citep[$208$ galaxies,][]{KASHINO17} and $z\sim2.3$ \citep[in total $360$ galaxies,][]{ERB06,GENZEL14,STEIDEL14,SHAPLEY15,STROM17}\footnote{Split into 155 galaxies from \citet{STEIDEL14}, 130 galaxies from \citet{SHAPLEY15}, and 75 galaxies from \citet{GENZEL14}.}. Similar to the SDSS samples, the stellar masses for these samples have been derived from a fit to the total galaxy photometry assuming constant and exponentially declining star formation histories. The photometry includes Spitzer imaging at $>2\,{\rm \mu m}$, which covers a similar rest-frame wavelength range as for the local galaxies. In the case of Kashino et al., stellar masses have been converted from a \citet{SALPETER55} to a Chabrier IMF.

    \subsection{Parameterization of the BPT main-sequence locus: $O3(N2,z)$}\label{sec:mainsequence}

	As outlined in Section~\ref{sec:idea} and indicated in Figure~\ref{fig:bpt2}, galaxies at a given redshift occupy a defined locus on the BPT diagram (the BPT main-sequence), similar to the $M-{\rm SFR}$ main-sequence. This locus shifts towards higher \oiiihb\ ratios at fixed \niiha\ (or, alternatively, higher \niiha\ at fixed \oiiihb) at high redshifts as shown by many spectroscopic studies \citep{ERB06,MASTERS14,STEIDEL14,SHAPLEY15,KASHINO17,STROM17}. This shift is attributed by \citet{MASTERS16} to an increasing ionization parameter and lower metallicity at fixed mass (and thus N/O ratio, which is mostly set by stellar mass as shown by the same study) of the average high-redshift galaxy. The evolution is also captured in compilations of the stellar mass vs. metallicity relation, the latter commonly estimated from these emission line ratios.
    
    In the following, we use the spectroscopic measurements at $z\sim1.6$ \citep{KASHINO17}\footnote{Note that the data from this study only covers the range $-1.2 < $ log(\niiha) $ < -0.1$ in contrast to the other samples used here. We therefore extrapolate the relation between \oiiihb\ and \niiha\ given by that study to reach log(\niiha) $= -1.6$. This extrapolation is well defined as the BPT relation narrows towards high \oiiihb\ and low \niiha. Furthermore, uncertainties in this extrapolation would only affect galaxies stellar masses $\logm \lesssim 8.5$, which is below what is considered in the following model.} and $z\sim2.3$ \citep{STEIDEL14,STROM17} to parameterize the BPT main-sequence as a function of redshift \citep[similar to][]{KEWLEY13}. We find that a simple shift in \niiha\ starting from the local locus \citep{KEWLEY08} is a good fit to the data (see also Figure~\ref{fig:bpt2}). We parameterize this shift proportional to cosmic time to obtain the following relation for the redshift dependence of the BPT main-sequence by a least-square fit
    
    \begin{equation}\label{eq:bptfit}
    O3(N2,z) = \frac{0.61}{N2 + \delta - \gamma\, (1 + z)^2} + 1.08,
    \end{equation}
     
     with $\delta=0.138\pm0.005$ and $\gamma=0.042\pm0.005$. Note that this parameterization might not be valid beyond $z\sim2.7$ as it cannot be tested with the current data at higher redshifts.
     Finally, we note that the $z\sim2.3$ sample by \citet{STEIDEL14} is the most comprehensive as it includes published stellar masses, \niiha, and \oiiihb~for all individual galaxies, which is crucial for our analysis. The BPT main-sequence has, however, been determined by other studies as well. In particular, we note here the sample at $z\sim2.3$ by \citet{SHAPLEY15} based on the MOSFIRE Deep Evolution Field (MOSDEF) survey, which BPT main-sequence is offset by $\sim-0.2\,{\rm dex}$ from the Steidel locus. Using their relation, we obtain for the redshift evolution (Equation~\ref{eq:bptfit}) parameters $\delta=0.110\pm0.005$ and $\gamma=0.032\pm0.005$. This defines an uncertainty due to sample biases and measurement differences of $0.028$ and $0.010$ in $\delta$ and $\gamma$, respectively.
     In Section~\ref{sec:scatter}, we discuss in detail the impact of sample biases on the parameterization of the BPT main-sequence and our final model.

   \subsection{Parameterization of $M(O3,N2)$}\label{sec:mass}

	Second, we parameterize the stellar mass distribution on a given BPT main-sequence locus. As argued in \citet{MASTERS16} and Section~\ref{sec:idea}, galaxies at high redshifts occupy a distinct region of the local BPT diagram, namely at higher \oiiihb\ for fixed \niiha. Importantly, they can still be described by the relations found in the SDSS data, although these data become sparse at the location of the high-redshift galaxies and therefore an extrapolation becomes necessary.
    In the following, we therefore parameterize the relation $M(O3,N2)$ observed in the local SDSS data to derive the stellar mass distribution by a cut through this plane along a given BPT locus. We find that the following functional form describes best the parabolic shaped (in O3) stellar mass isochrones that are displaced in N2
\begin{equation}\label{eq:mass}
M(O3,N2) = A + B\, (N2 + \alpha)  + C\, (O3 + \beta)^2,\\
\end{equation}

where $A$, $B$, $C$, $\alpha$, and $\beta$ are determined by a Levenberg$-$Marquardt algorithm, as part of the \texttt{R/minpack.lm} package\footnote{\url{https://cran.r- project.org/web/packages/minpack.lm/index.html}} \citep{minpack_lm}.
The fitting to the SDSS data is performed on medians derived from a binning in \niiha~and \oiiihb\ to increase the S/N in the data (e.g., upper panel in Figure~\ref{fig:bpt2}). We set the weights for the fitting proportional to the number of galaxies per bin (chosen to be $15$ or more). We do not fit galaxies below \loiiihb~$=-0.6$ that correspond to the most massive galaxies in SDSS and might include post-starburst galaxies and almost quiescent galaxies that could bias our fit.
	Figure~\ref{fig:bptfit} shows the data (\textit{left}) together with the best-fit model (\textit{middle}) and residual (\textit{right}). The best-fit parameters for Equation~\ref{eq:mass} are $A = 7.689 \pm 0.450$, $B = 3.696 \pm 0.005$, $C = 1.960 \pm 0.015$, $\alpha = 1.126 \pm 0.060$, and $\beta = 0.273 \pm 0.005$.
	Our best fit describes the stellar masses to better than $0.1\,{\rm dex}$ for most values of \oiiihb~and \niiha. It under-predicts the stellar mass for galaxies on the lower crest of the local BPT main-sequence by up to $0.3\,{\rm dex}$. This is mostly an effect of the weighting, which is chosen to minimize the residual at the local BPT main-sequence as well as on the upper crest of the BPT locus where galaxies with higher sSFR (or redshift) are located. 
	
\begin{figure}
\centering
\includegraphics[width=1.0\columnwidth, angle=0]{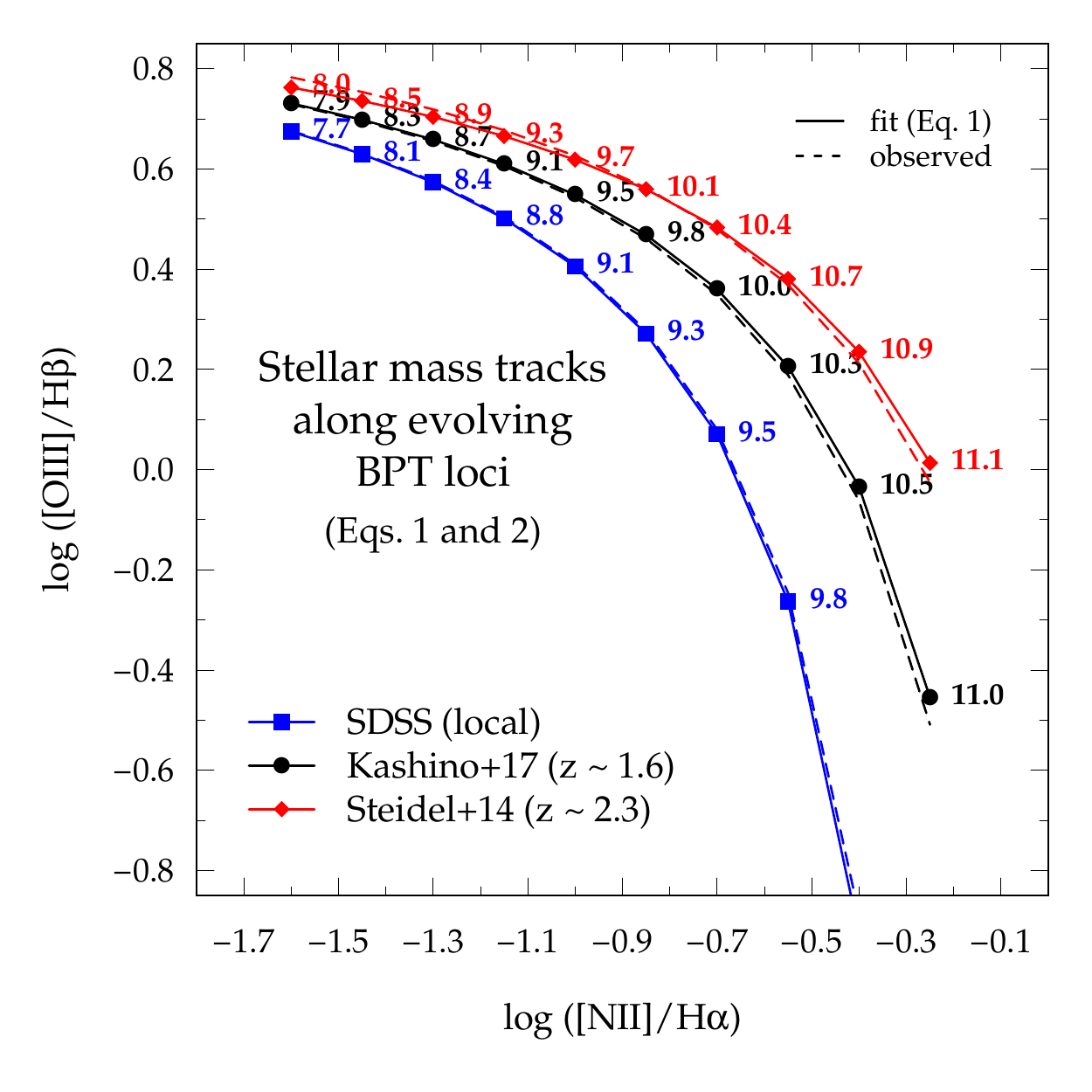}
\caption{Stellar mass tracks on the BPT diagram for fixed BPT loci at redshifts of $z\sim0$ \citep{KEWLEY08}, $z\sim1.6$ \citep{KASHINO17}, and $z\sim2.3$ \citep{STEIDEL14}. These three data sets are used to derive the shift of the BPT locus as a function of redshift (solid lines, Equation~\ref{eq:bptfit}). The observed loci at $z\sim0$, $z\sim1.6$, and $z\sim2.3$ are shown as dashed lines for comparison. The stellar mass tracks, $M(N2,z)$, are parameterized in the final Equation~\ref{eq:final}, with stellar masses given as $\logm$.
\label{fig:masstracks}}
\end{figure}

    In Figure~\ref{fig:masstracks} we show the BPT main-sequence loci at $z\sim0$, $z\sim1.6$, and $z\sim2.3$ from Equation~\ref{eq:bptfit} with indicated stellar masses from Equation~\ref{eq:mass}.
    Note that at a fixed stellar mass the \niiha\ ratio decreases and the \oiiihb\ ratio increases with redshift, which is essentially the stellar mass vs. metallicity relation changing across cosmic time as quoted by many studies in the literature \citep[e.g.,][]{MAIOLINO08,LILLY13}\footnote{Recall that the gas-phase metallicity, $\oabund$, is inversely proportional to \oiiihb\ and proportional to \niiha\ \citep[e.g.,][]{MAIOLINO08}.}. In this sense, the above equations also give an empirical parameterization of the evolution of the mass$-$metallicity relation with redshift.

\begin{figure}
\centering
\includegraphics[width=1.0\columnwidth, angle=0]{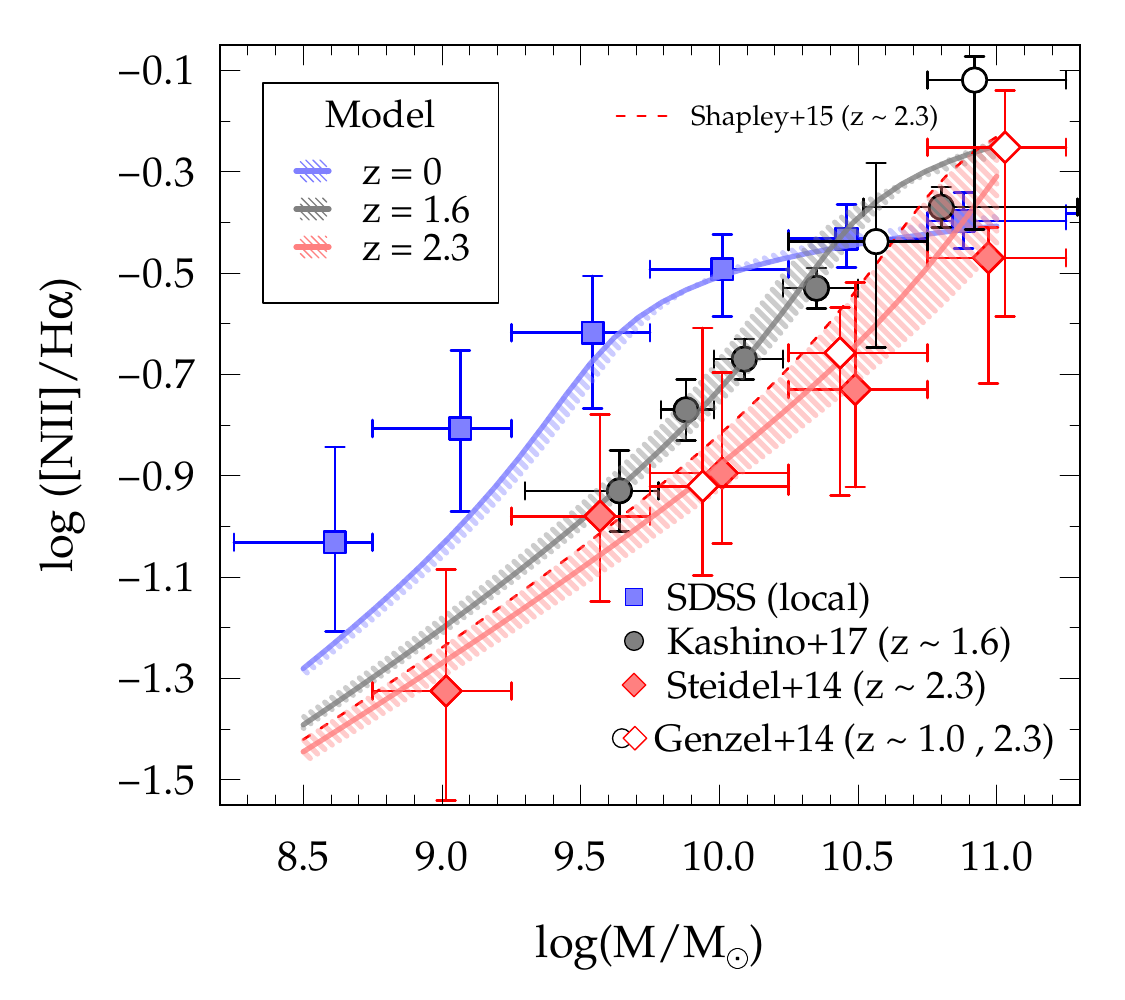}
\caption{Relation between stellar mass and \niiha\ ratio from observed data at $z\sim0$ (blue squares), $z\sim1.6$ (black points), and $z\sim2.3$ (red diamonds) from SDSS, \citet{KASHINO17}, and \citet{STEIDEL14}, respectively. We also show data from \citet{GENZEL14} at $z\sim1$ and $z\sim2.3$ with open symbols for reference (see Section~\ref{sec:scatter} for discussion). Our model predictions (Equation~\ref{eq:final}) at the median redshift of the samples are shown as lines and the hatched regions show the range in model \niiha\ values for the redshift distribution of the observations. Our parameterization reproduces the data within $\sim0.1\,{\rm dex}$, well within its scatter. The dashed red line indicates the $z\sim2.3$ relation from MOSDEF \citep{SHAPLEY15} with slightly larger \niiha\ ratios at a given mass compared to the \citet{STEIDEL14} relation at the same redshift. This difference can be explained by different selections of the two samples (see Section~\ref{sec:scatter}).
\label{fig:n2prediction}}
\end{figure}

\begin{figure}
\centering
\includegraphics[width=1.0\columnwidth, angle=0]{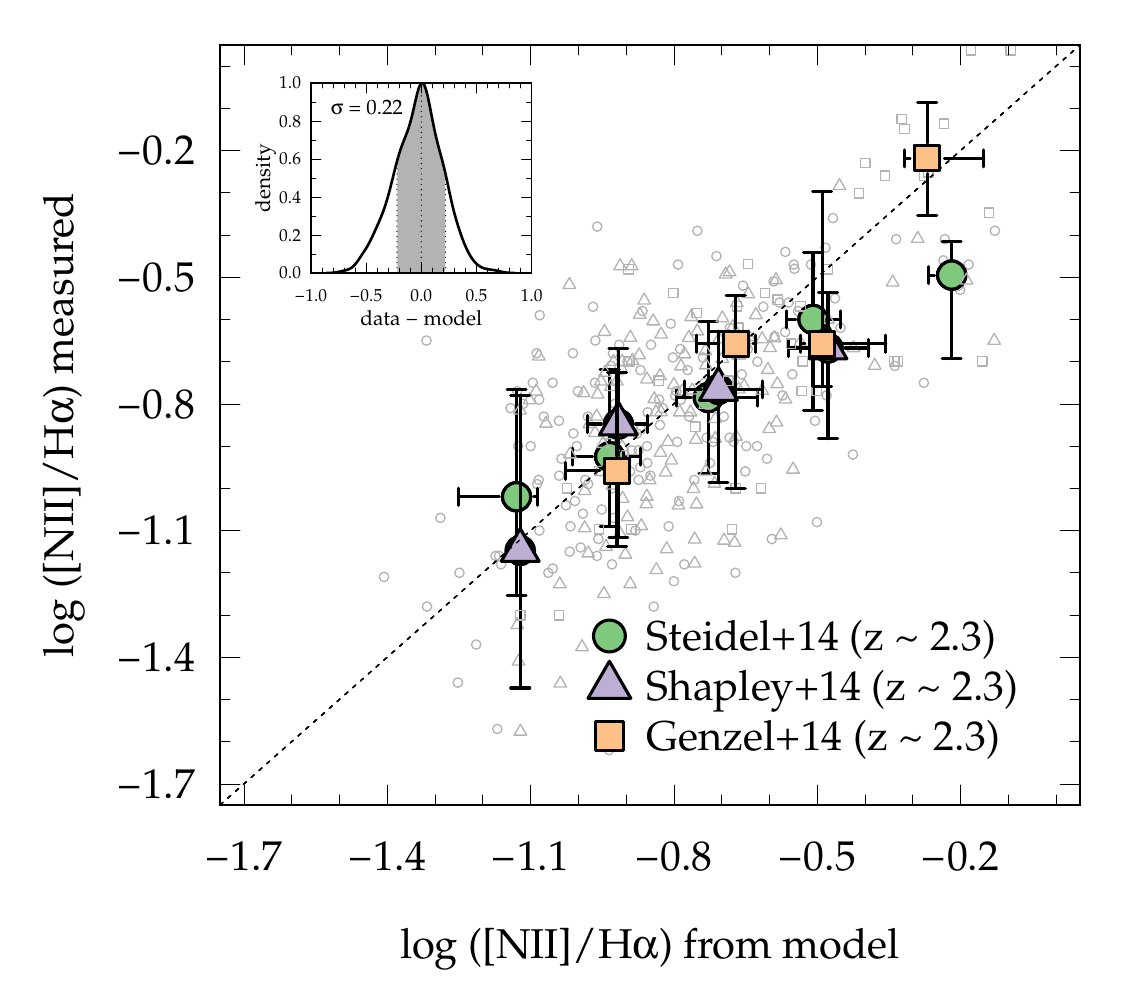}
\caption{Quantification of scatter in the predicted \niiha\ ratios based on the \citet{STEIDEL14} (circles), \citet{SHAPLEY15} (triangles), and \citet{GENZEL14} (squares) samples at $z\sim2.3$. The individual measurements are shown in gray, the medians with $1\sigma$ scatter from data and model are shown as large filled symbols in green, purple, and orange, respectively. The 1-to-1 relation is indicated by the dotted line. For our model predictions of \niiha, we expect a scatter of $\sim0.22\,{\rm dex}$ (see inset), which we find to be constant with stellar mass.
\label{fig:n2scatter}}
\end{figure}

     \subsection{Final N2(M,z) parameterization}\label{sec:final}
     
     The combination of Equations~\ref{eq:bptfit} and \ref{eq:mass}, allows us to parameterize the \niiha\ ratio as a function of stellar mass and redshift

\begin{multline*}
 M(N2,z) = 3.696\,\xi + 3.236\,\xi^{-1} + 0.729\,\xi^{-2}\\
 + 14.928 + 0.156\,(1+z)^2,
\end{multline*}

     with
     
     \begin{equation}\label{eq:final}
     \xi(N2,z) \equiv N2 + 0.138 - 0.042\,(1+z)^2.
     \end{equation}
     
     This equation can be reversed numerically to obtain $N2(M,z)$. For visual clarity, we do not reverse this equation algebraically, but we provide Table~\ref{tab:lookup} for a convenient lookup of $N2(M,z)$.
     
     In the following, we test our model on data at $z\sim0$, $z\sim1.6$, and $z\sim2.3$. Figure~\ref{fig:n2prediction} shows the relation between stellar mass and \niiha\ at the three different redshifts. For $z\sim0$ and $z\sim2.3$, the symbols show the medians in stellar mass and \niiha\ with $1\sigma$ scatter from SDSS and \citet{STEIDEL14}, for $z\sim1.6$ we use the median stacks in stellar mass provided by \citet{KASHINO17}, thus the errors represent the error on the median and not the actual scatter. We also show the medians at $z\sim1$ and $z\sim2.3$ from \citet{GENZEL14} covering the massive end of the galaxy mass function.
     Our parameterization at $z=0$, $1.6$, and $2.3$ using Equation~\ref{eq:final} is shown with lines. The hatched regions show the range of our model values for the redshift distribution of the observed samples.
    Our model predicts the \niiha\ ratio in general within $\sim0.1\,{\rm dex}$ of the observed data, which is well within the scatter of the data at all redshifts $0 \lesssim z \lesssim 2.3$ and stellar masses $9.5 < \logm < 11.0$. This is remarkable since the relation between \niiha\ and stellar mass is solely based on the local SDSS data without information from higher redshifts.
     However, we note that our parameterization systematically under-predicts the \niiha\ ratios at very low masses ($\logm \lesssim 9.5$) and low redshifts ($z \sim 0$) by up to $0.2\,{\rm dex}$. Furthermore, we notice that our model over-predicts \niiha\ ratios of the most massive galaxies ($\logm \sim 11.0$) in the \citet{STEIDEL14} sample systematically by $\sim0.1\,{\rm dex}$.
     The former is likely due to the generally larger residuals in the parameterization of $M(O3,N2)$ for local low-mass galaxies (right panel of Figure~\ref{fig:bptfit}).
     The latter can be explained two-fold. First, Equation~\ref{eq:mass} is an extrapolation at $\logm \gtrsim 11.0$ as there are very few star-forming galaxies that are massive and low metallicity in the local SDSS sample. Specifically, there are only $191$ star-forming galaxies at $\logm > 10.8$ with \oiiihb\ $>1$ (the region on the BPT diagram that is occupied by the high-redshift samples), which represents less than 0.1\% of the total sample. At $\logm > 11$ this amount reduces to $73$ galaxies. Second, the statistics of massive high-redshift galaxies in current spectroscopic samples is poor and dominated by sample selection and cosmic variance. Especially, we note that our model predicts almost perfectly the \niiha\ flux ratios of the sample by \citet{GENZEL14} who specifically targeted massive galaxies at $z\sim2$ (confirmed AGNs removed).

     \subsection{Scatter in \niiha~line ratios}\label{sec:scatter}
     
     Our model provides \emph{median} \niiha\ ratios for a given redshift and stellar mass. This median is mainly defined by the BPT main-sequence locus that we parameterized in Section~\ref{sec:mainsequence}. Deviations from this locus will lead to a physical scatter around the median \niiha~ratios provided by our model.  Here, we study the origin and amplitude of this scatter in more detail as well as the impact of measurement uncertainties.

     Figure~\ref{fig:n2scatter} compares the true \niiha~values to the ones obtained from our model at $z\sim2.3$ based on the \citet{STEIDEL14}, \citet{GENZEL14}, and \citet{SHAPLEY15}\footnote{Stellar masses and \niiha~measurements are taken from \citet{SANDERS17}. No redshifts are published for individual galaxies, therefore we assume $z=2.3$ for all galaxies.} samples, for which these measurements (\niiha~and stellar mass) are published for individual galaxies. Apart from the good agreement on average between model and true \niiha~values, we measure a (log-symmetric) $1\sigma$ scatter of $0.22\,{\rm dex}$ (inset in Figure~\ref{fig:n2scatter}), which we find to be constant with \niiha\ ratio (hence stellar mass). This scatter is identical for the individual samples at $z\sim2.3$. The same computation for $z\sim1.6$ and local galaxies reveal a scatter of $0.21\,{\rm dex}$ and $0.13\,{\rm dex}$, respectively (Appendix~\ref{app:scatter}).
     This scatter is introduced by differences in physical properties of the galaxies as well as measurement uncertainties as discussed below.
 
     \subsubsection{Physical scatter due to sSFR}
     The upper panel in  Figure~\ref{fig:bpt2} shows that mainly stellar mass determines the position of galaxies on the BPT diagram for a given BPT main-sequence as parameterized in Section~\ref{sec:mainsequence} as a function of redshift. On the other hand (as shown on the lower panel of Figure~\ref{fig:bpt2}), the sSFR varies mostly \emph{perpendicular} to the BPT main-sequence loci. We therefore argue that sSFR acts as a secondary parameter defining the location of galaxies on the BPT diagram at a fixed stellar mass and redshift.
     By fixing a BPT main-sequence for our model, we indirectly assume a median sSFR given by the sample that is used to anchor our model (in our case the average sSFR of the \citet{STEIDEL14} sample, which represents well the stellar mass vs. SFR main-sequence at $z\sim2.3$). 
      Because of the remarkably constant $\sim0.3\,{\rm dex}$ scatter of the stellar mass versus SFR relation \citep[e.g.,][]{DADDI07a,NOESKE07,SCHREIBER15,TOMCZAK16}, galaxies at a fixed stellar mass and redshift show a range in sSFR, hence inducing a scatter perpendicular to the average BPT main-sequence locus.
     This (physical) scatter directly translates into the scatter seen in our comparison of true and model \niiha~ratios.
     
     The effect of selection biases on the BPT main-sequence can be seen by comparing the result of the MOSDEF \citep{SHAPLEY15} and \citet{STEIDEL14} studies.
     The BPT main-sequence locus at $z\sim2.3$ derived from the MOSDEF survey is offset by up to $\sim -0.2\,{\rm dex}$ in \niiha\ from the \citet{STEIDEL14} locus, although both samples have almost identical distribution in stellar mass and redshift. Using the MOSDEF locus for our model would therefore result in up to $0.1-0.2\,{\rm dex}$ larger \niiha\ flux ratios at a fixed stellar mass (dashed line in Figure~\ref{fig:n2prediction}).
    The physical reason for the seeming discrepancy is likely a slight excess of high sSFR galaxies in the \citet{STEIDEL14} sample compared to the MOSDEF sample \citep[as also pointed out by][]{SHAPLEY15}, in agreement with our identification of sSFR as a secondary parameter. This excess in sSFR could be caused by the UV color selection in the case of the \citet{STEIDEL14} sample, which favors higher star formation compared to a continuum or stellar mass selected sample as in the case of MOSDEF.
    
    	An ``emission line complete'' sample would allow us to derive the correct average BPT main-sequence locus at a given redshift and hence anchor our model at high redshifts, however, selecting such a sample is almost impossible at these redshifts since there will always be certain selection biases. At $z\sim2.3$, the likely average locus would be somewhere in between the MOSDEF and \citet{STEIDEL14} derivations and therefore not far from our model predictions (see Figure~\ref{fig:n2prediction}).
    
    \subsubsection{Scatter due to measurement uncertainties}
    
    In addition to differences in stellar mass and sSFR, measurement uncertainties can contribute to the scatter on the BPT diagram for a given sample at a given redshift. This is becoming increasingly more valid at higher redshift where the measurements become lower S/N. These concerns can dilute a clear BPT main-sequence locus.
    
    \citet{STEIDEL14} quote an intrinsic scatter (i.e., corrected for measurement uncertainties) of $0.12\,{\rm dex}$ on the BPT main-sequence locus (in \niiha\ and \oiiihb) at $z\sim2.3$. This is consistent with the measurements by \citet{SHAPLEY15} at the same redshift, \citet{KASHINO17} at $z\sim1.6$, and for local galaxy samples \citep[$\sim 0.11\,{\rm dex}$,][]{KEWLEY08}.

	We measure an observed scatter of $0.22\,{\rm dex}$, $0.21\,{\rm dex}$ and $0.13\,{\rm dex}$ between true and model \niiha\ ratios at $z\sim 2.3$, $z\sim1.6$ and $z\sim0$, respectively (see also Appendix~\ref{app:scatter}). Comparing this to the intrinsic scatter in the BPT main-sequence loci given above suggests that roughly half of the uncertainties in the model derived \niiha\ ratios at $z\sim1.6$ and $2.3$ are due to the combined uncertainties in the individual measurements and our model.

\begin{deluxetable*}{c cccccccccccccc}
\tabletypesize{\scriptsize}
\tablecaption{Look-up table for \niiniihatot\ flux ratios (including both \nii\ emission lines) given in linear scale from $8.5 < \logm < 11.1$ and $0 < z < 2.6$ derived by Equation~\ref{eq:final}$^\dagger$. These values can be used for the conversion of the observed \niiphatot\ flux to the intrinsic \halpha\ flux.\label{tab:lookup}}
\tablewidth{0pt}
\tablehead{
\multicolumn{1}{c}{$\logm$} & \multicolumn{14}{c}{Redshift} \\[-0.2cm]
\multicolumn{1}{c}{------------------} & \multicolumn{14}{c}{---------------------------------------------------------------------------------------------------------------------------} \\[-0.2cm]
\colhead{} & \colhead{$0.0$} & \colhead{$0.2$} & \colhead{$0.4$} & \colhead{$0.6$} & \colhead{$0.8$} & \colhead{$1.0$} & \colhead{$1.2$} & \colhead{$1.4$} & \colhead{$1.6$} & \colhead{$1.8$} & \colhead{$2.0$} & \colhead{$2.2$} & \colhead{$2.4$} & \colhead{$2.6$} 
}
\startdata
$8.5$ & $0.07$ & $0.06$ & $0.06$ & $0.06$ & $0.06$ & $0.06$ & $0.05$ & $0.05$ & $0.05$ & $0.05$ & $0.05$ & $0.05$ & $0.04$ & $0.04$ \\ 
$8.7$ & $0.08$ & $0.08$ & $0.08$ & $0.07$ & $0.07$ & $0.07$ & $0.07$ & $0.06$ & $0.06$ & $0.06$ & $0.06$ & $0.05$ & $0.05$ & $0.05$ \\ 
$8.9$ & $0.10$ & $0.10$ & $0.09$ & $0.09$ & $0.08$ & $0.08$ & $0.08$ & $0.07$ & $0.07$ & $0.07$ & $0.07$ & $0.06$ & $0.06$ & $0.06$ \\ 
$9.1$ & $0.12$ & $0.12$ & $0.11$ & $0.11$ & $0.10$ & $0.10$ & $0.09$ & $0.09$ & $0.08$ & $0.08$ & $0.08$ & $0.07$ & $0.07$ & $0.07$ \\ 
$9.3$ & $0.16$ & $0.15$ & $0.14$ & $0.13$ & $0.13$ & $0.12$ & $0.11$ & $0.11$ & $0.10$ & $0.10$ & $0.09$ & $0.09$ & $0.08$ & $0.08$ \\ 
$9.5$ & $0.21$ & $0.20$ & $0.19$ & $0.17$ & $0.16$ & $0.15$ & $0.14$ & $0.13$ & $0.12$ & $0.11$ & $0.11$ & $0.10$ & $0.10$ & $0.09$ \\ 
$9.7$ & $0.26$ & $0.25$ & $0.24$ & $0.23$ & $0.21$ & $0.19$ & $0.17$ & $0.16$ & $0.15$ & $0.14$ & $0.13$ & $0.12$ & $0.11$ & $0.11$ \\ 
$9.9$ & $0.28$ & $0.28$ & $0.28$ & $0.28$ & $0.26$ & $0.24$ & $0.22$ & $0.20$ & $0.18$ & $0.16$ & $0.15$ & $0.14$ & $0.13$ & $0.13$ \\ 
$10.1$ & $0.30$ & $0.30$ & $0.31$ & $0.31$ & $0.31$ & $0.30$ & $0.28$ & $0.25$ & $0.22$ & $0.20$ & $0.18$ & $0.17$ & $0.16$ & $0.15$ \\ 
$10.3$ & $0.31$ & $0.32$ & $0.32$ & $0.33$ & $0.33$ & $0.34$ & $0.33$ & $0.32$ & $0.29$ & $0.25$ & $0.22$ & $0.20$ & $0.19$ & $0.17$ \\ 
$10.5$ & $0.32$ & $0.33$ & $0.34$ & $0.34$ & $0.35$ & $0.36$ & $0.36$ & $0.36$ & $0.35$ & $0.32$ & $0.28$ & $0.25$ & $0.22$ & $0.21$ \\ 
$10.7$ & $0.33$ & $0.34$ & $0.35$ & $0.36$ & $0.36$ & $0.37$ & $0.38$ & $0.39$ & $0.39$ & $0.38$ & $0.35$ & $0.31$ & $0.27$ & $0.24$ \\ 
$10.9$ & $0.34$ & $0.35$ & $0.35$ & $0.36$ & $0.37$ & $0.39$ & $0.40$ & $0.41$ & $0.42$ & $0.42$ & $0.42$ & $0.38$ & $0.33$ & $0.29$ \\ 
$11.1$ & $0.34$ & $0.35$ & $0.36$ & $0.37$ & $0.38$ & $0.40$ & $0.41$ & $0.42$ & $0.44$ & $0.45$ & $0.46$ & $0.45$ & $0.41$ & $0.36$ \\ 
\enddata
\tablenotetext{\dagger}{Note that Equation~\ref{eq:final} is basically a parameterization of the stellar mass vs. gas-phase metallicity relation as a function of redshift.}
\end{deluxetable*}

\subsection{Large \niiha\ ratios in massive high-z galaxies}\label{sec:agns}
	Our model predicts larger \niiha\ ratios in massive ($\logm\sim11.0$) high-redshift galaxies compared to similar massive galaxies at $z\sim0$ (Figure~\ref{fig:n2prediction}). This is also suggested by the shifted high-redshift BPT loci to higher \oiiihb\ and \niiha\ ratios (Figure~\ref{fig:masstracks}).
	This could be caused by an increasing amount of galaxies with broad line emission at such high stellar masses.  
     In fact, \citet{GENZEL14} study the statistics of broad \halpha\ and \nii\ emission in samples of star-forming galaxies at $1 < z < 3$ and suggest broad nuclear components due to a combination of shocks and photoionization and also AGNs in more than half of these galaxies at $\logm \sim 11.0$. Specifically, they suggest that the contribution of such galaxies to high-mass samples is at least as large as AGN samples selected with X-ray, optical, infrared, or radio indicators. The median \niiha\ ratio per stellar mass bin from the \citet{GENZEL14} sample is higher than the one of the \citet{STEIDEL14} sample and in good agreement with our model (Figure~\ref{fig:n2prediction}).
     Along the same lines, \citet{KEWLEY13} motivates a shift of the separation line between normal star-forming galaxies and AGNs to higher \niiha\ and \oiiihb\ ratios by means of increased photoionization at higher redshifts.

\section{Implications}\label{sec:implications}

	In the previous section, we have derived an empirical parameterization to predict the \niiha\ flux ratios for galaxies up to $z\sim3$ and $\logm = 11.0$. Here, we study in detail the implications of our model on
    \begin{itemize}
    \item Data interpretation of current low spectral resolution surveys (Section~\ref{sec:datainterp}),
    \item The \nii\ contamination of future flux-limited surveys (Section~\ref{sec:n2contamination}),
    \item Spectroscopic redshift measurements from the blended \halpha\ and \nii\ lines (Section~\ref{sec:centroid}),
    \item Expected \halpha\ emitter number counts of future surveys derived from current \halpha\ LFs (Section~\ref{sec:half}). 
    \end{itemize}

    For surveys with low spectral resolution, \halpha\ is commonly blended with \textit{both} \nii\ emission lines (at rest-frame $6548\,{\rm \AA}$ and $6584\,{\rm \AA}$). Hence, a more useful quantity to quote is the \emph{total \nii\ flux contamination fraction}, which we define as (\nii$\lambda\lambda6548,6584)/($\nii$\lambda\lambda6548,6584+$\halpha), in short \niiniihatot.
    In the following we assume for the flux of the second \nii\ line, blue-ward of \halpha, \nii$\lambda6548=\frac{1}{3}$\nii$\lambda6584$ \citep{ACKER89}. In Table~\ref{tab:lookup} we provide \niiniihatot\ flux ratios in linear scale as a function of stellar mass and redshift derived from our Equation~\ref{eq:final}. This table can be used as a convenient tool to convert observed \niiphatot\ fluxes and luminosities (including both \nii) into intrinsic \halpha\ fluxes and luminosities.

    \subsection{Data interpretation of current surveys at low spectral resolution}\label{sec:datainterp}
    
    In current grism surveys with low spectral resolution where \nii\ and \halpha\ are not resolved, such as WISPS, a constant total \nii\ flux contamination fraction is commonly used to obtain intrinsic \halpha\ values from which SFRs or \halpha\ luminosity functions are measured \citep{COLBERT13,MEHTA15}. The value generally applied is $0.29$ (i.e., $29\%$, or $F_{\rm H\alpha}= 2.5 \times F_{\rm [NII]}$ that follows from \niiniihatot $=0.29$) according to the average population of galaxies at $z\sim0$ with an \halpha\ EW of less than $200\,{\rm \AA}$. However, such an assumption can be misleading, since the true \niiha\ flux ratio can vary by an order of magnitude across samples depending on redshift and stellar mass (Figure~\ref{fig:n2prediction}).
    Other studies apply a variable \nii\ contamination correction using the relation between \niiha\ and the \halpha\ equivalent-width ($\propto$ sSFR), resulting in a slightly lower median \nii\ contamination of $\sim 24\%$ \citep[e.g.,][]{VILLAR08,SOBRAL09,LEEJANICE12,SOBRAL12,SOBRAL13}. While this approach is more accurate, it could still miss the dependency with stellar mass.
        
\begin{figure}
\centering
\includegraphics[width=1.0\columnwidth, angle=0]{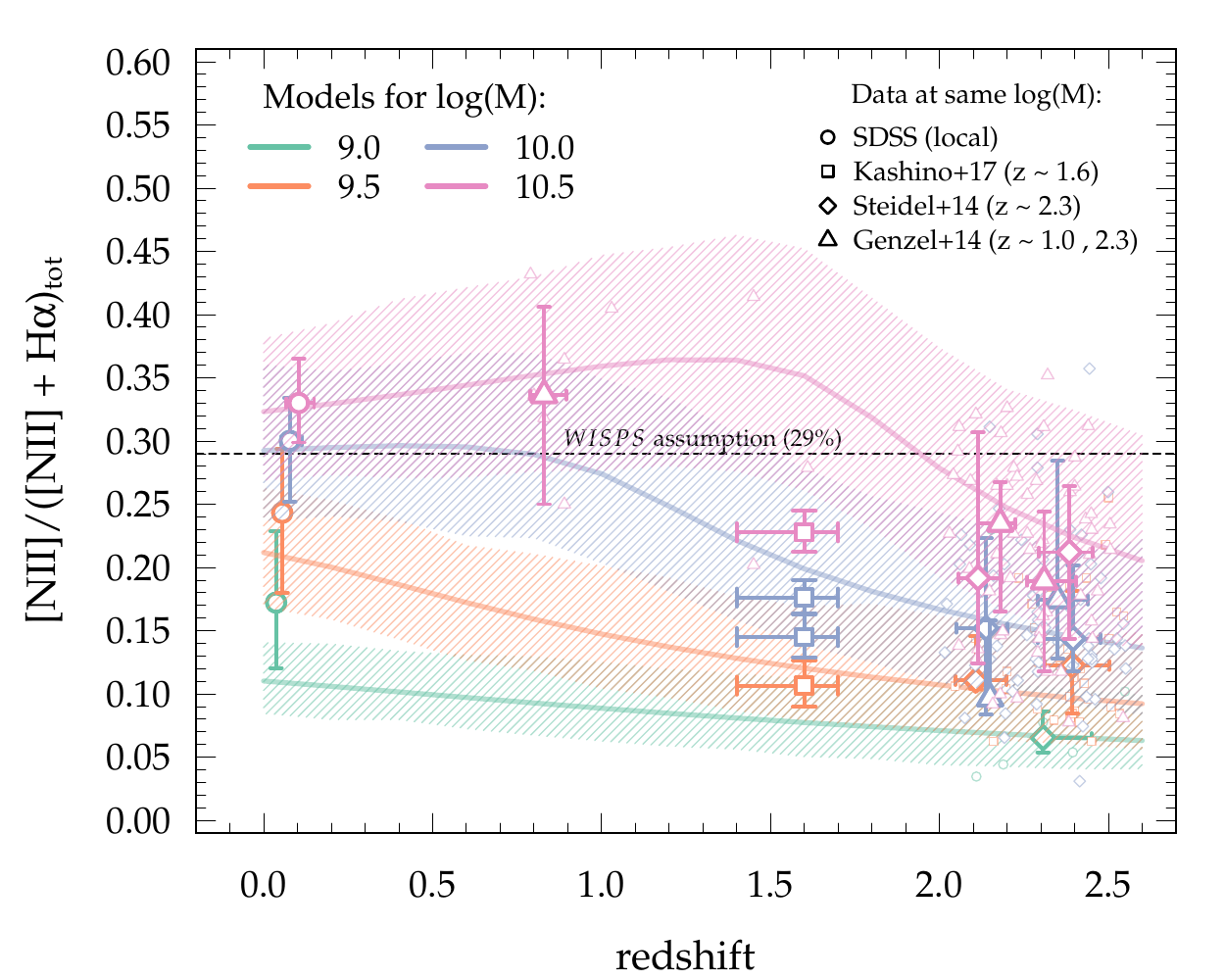}\\
\caption{Total \nii\ flux contamination fraction as a function of redshift for 4 different stellar masses from our model (lines). Medians of observations are shown with large symbols of the same colors, individual measurements are shown with small symbols. The WISPS assumption \citep{COLBERT13} of a constant $29\%$ \nii\ contamination is shown as dashed line for reference. Note that this assumption is only justified for $\logm\sim10.0$ galaxies at $z < 1$ and $\logm\sim10.5$ galaxies at $z\sim2$ and otherwise over- or underestimates the true contamination by a significant factor.
\label{fig:n2massz}}
\end{figure}

    To study the accuracy of a constant correction, we show in Figure~\ref{fig:n2massz} the total \nii\ flux contamination fraction on linear scaling as a function of redshift at four different stellar masses. Our model is shown as lines with the hatched region corresponding to the approximate, redshift dependent scatter (see Section~\ref{sec:scatter}). Observed data at $0 < z < 2.3$ binned in redshift and stellar mass ($\Delta\logm = 0.5$ around stellar masses shown) is shown with symbols and for reference a constant total \nii\ flux contamination fraction of $29\%$ is indicated by the horizontal dashed line.
    We find that the assumption of a constant contamination of $29\%$ is only justified for galaxies at $z < 1$ and $\logm\sim10.0$ as well as $z\sim2$ and $\logm\sim10.5$. Otherwise, our model shows that such an assumption generally overestimates the true contamination, which can be as severe as a factor of $3$ for $\logm \sim 9$ galaxies at all redshifts. But also the \nii\ contamination of high-mass ($\logm > 10.0$) galaxies at $z\sim1.5-2$ is overestimated by factors of $1.5-2$.
    
    This has important consequences for the intrinsic \halpha\ measurements by WISPS. Specifically, since this sample spans a range in stellar mass of $8.5 < \logm < 10$ at $0.5 < z < 1.5$ \citep{ATEK10,HENRY13}, we expect the \halpha\ luminosities to be systematically underestimated by $20-30\%$ or more on average. Furthermore, since \niiha\ is a function of stellar mass and therefore (via the SFR) also a function of \halpha\ luminosity, the shape of the \halpha\ LF is affected, which has an impact on number counts for future surveys (see Section~\ref{sec:half}).

\begin{figure}
\centering
\includegraphics[width=1.00\columnwidth, angle=0]{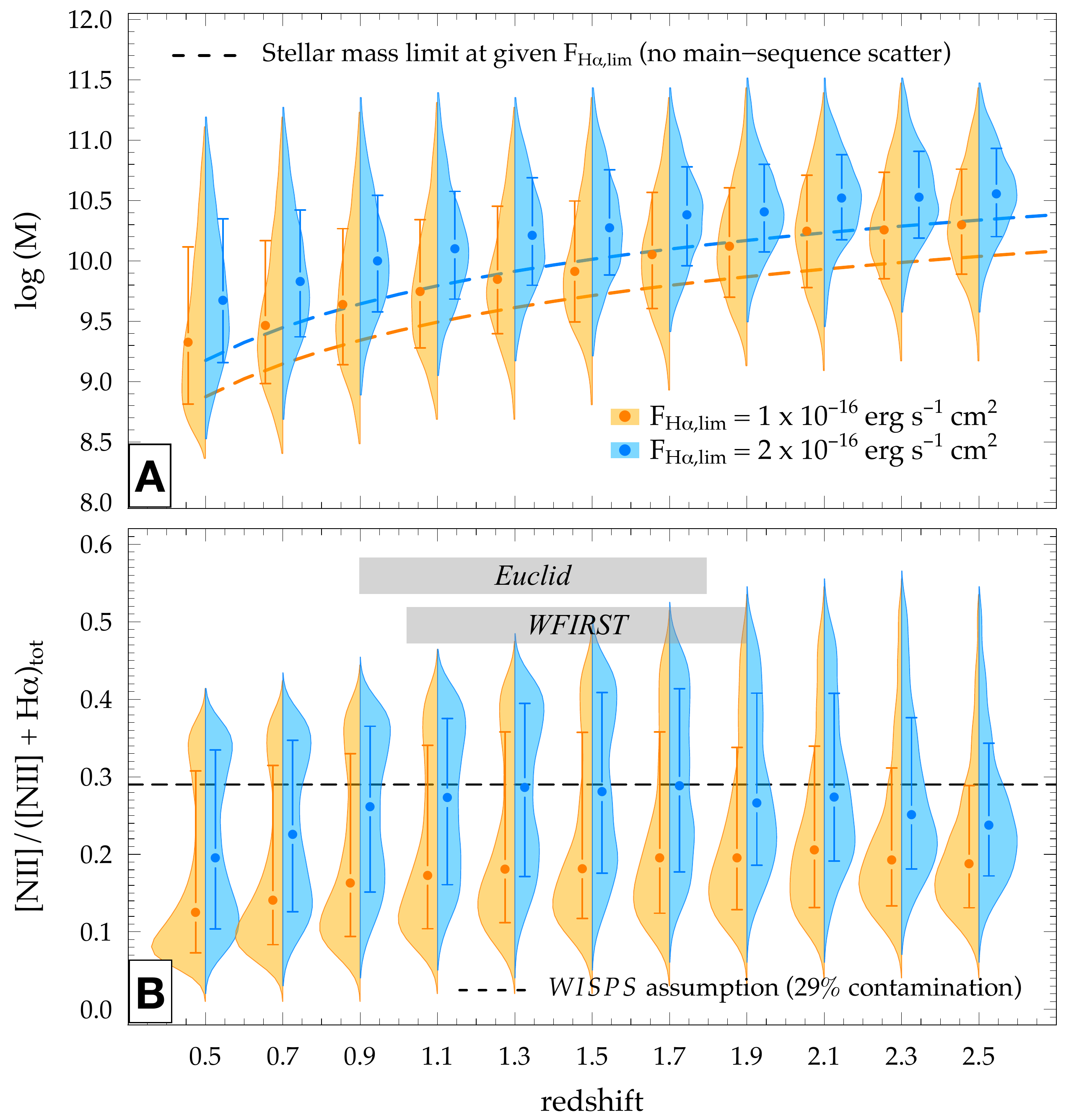}\\
\caption{Expected stellar mass \textit{(A)} and total \nii\ flux contamination fraction \textit{(B)} distributions as a function of redshift for flux limited surveys at $1\times 10^{-16}\,\ergPerSecondPerCM$ (orange) and $2\times 10^{-16}\,\ergPerSecondPerCM$ (blue), similar to expectations for \emph{WFIRST} and \emph{Euclid}, respectively. The medians of the distributions (points with error bars) are slightly displaced in redshift for clarity.  Our prediction is based on the empirical stellar mass vs. SFR relation (see text for details on derivation). For a survey like \emph{Euclid} ($0.9 < z < 1.8$) or \emph{WFIRST} ($1 < z < 2$), we expect a large variation in the total \nii\ flux contamination fraction of $5-40\%$ ($z=1$) and $10-45\%$ ($z=1.8$).
\label{fig:n2sim_all}}
\end{figure}

    \subsection{Total \nii\ flux contamination fraction distribution for future flux limited surveys}\label{sec:n2contamination}
    
    Future large area surveys at low spectral resolution such as \emph{Euclid} will suffer from \nii\ and \halpha\ blending.
    A proper de-blending of these lines is important for measuring physical quantities from \halpha\ such as SFR, galaxy kinematics, or dark matter properties, but also for accurate spectroscopic redshift used for cosmology.
    Here, we present realistic zeroth-order predictions for the total \nii\ flux contamination fraction as a function of redshift for flux-limited surveys at $0.5 < z < 2.5$.
    In the following, we assume observed line flux limits of $1\times10^{-16}\,\ergPerSecondPerCM$ and $2\times 10^{-16}\,\ergPerSecondPerCM$, similar to expectations for \emph{WFIRST} \citep[$5\sigma$ for a source of radius $0.3\arcsec$\footnote{
    \emph{WFIRST} Formulation Science Working Group, 2017, private communication.}, see also][]{SPERGEL15} and \emph{Euclid} \citep[$3.5\sigma$ for a source with diameter $0.6\arcsec$,][]{VAVREK16}, around the observed wavelength of \halpha.
    
	From the observed redshift-dependent stellar mass functions \citep{ILBERT13,DAVIDZON17}, we draw $10,000$ galaxies, to which we assign a SFR via the observed relation between stellar mass and SFR (main-sequence of star-forming galaxies) as parameterized by \citet{SCHREIBER15} including a scatter of $0.3\,{\rm dex}$. We then select galaxies above a SFR threshold derived from the line flux limits, which we converted to limiting \halpha\ luminosities (at given redshift) and then SFRs using the \citet{KENNICUTT98} description. The resulting stellar mass distributions for the two flux limits as a function of redshift are shown on the left (orange) and right (blue) side, respectively, of the ``Violin diagram'' in panel \textit{(A)} of Figure~\ref{fig:n2sim_all}. Note that the distributions extend across the sharp stellar mass limit derived from the observed line flux limits (dashed lines) because of the scatter of the star-forming main-sequence.

    Panel \textit{(B)} shows the corresponding distributions of the total \nii\ flux contamination fraction as a function of redshift.  The expected range in \niiniihatot\ is large because of the wide distribution in the stellar masses. Furthermore it is important to note that the distribution is double-peaked out to $z\sim1.5$. The first peak is due to the dominant number of low-mass galaxies (with low \niiha\ ratios), while the second peak arises due to the flattening of the stellar mass vs. \niiha\ ratio relation at large stellar masses. This can be seen clearly in Figure~\ref{fig:n2prediction} where the \niiha$-$M relation at $z=0$ flattens for stellar masses above approximately $\logm \sim 10$.
    Towards higher redshifts, the stellar mass distribution becomes tighter and the \nii\ contamination becomes single-peaked because of the increasing luminosity limit. 

	For a \emph{Euclid}-like survey ($0.9 < z < 1.8$), we expect a large variation in the total \nii\ flux contamination fraction of $10-40\%$ ($z=1$) and $15-45\%$ ($z=1.8$). For a \emph{WFIRST}-like survey ($1.0 < z < 1.9$), these numbers are lower ($5-40\%$ and $10-45\%$, respectively) due to the higher line sensitivity allowing to probe more galaxies at lower stellar masses, hence lower \niiha\ ratios.

\begin{figure}
\centering
\includegraphics[width=1.00\columnwidth, angle=0]{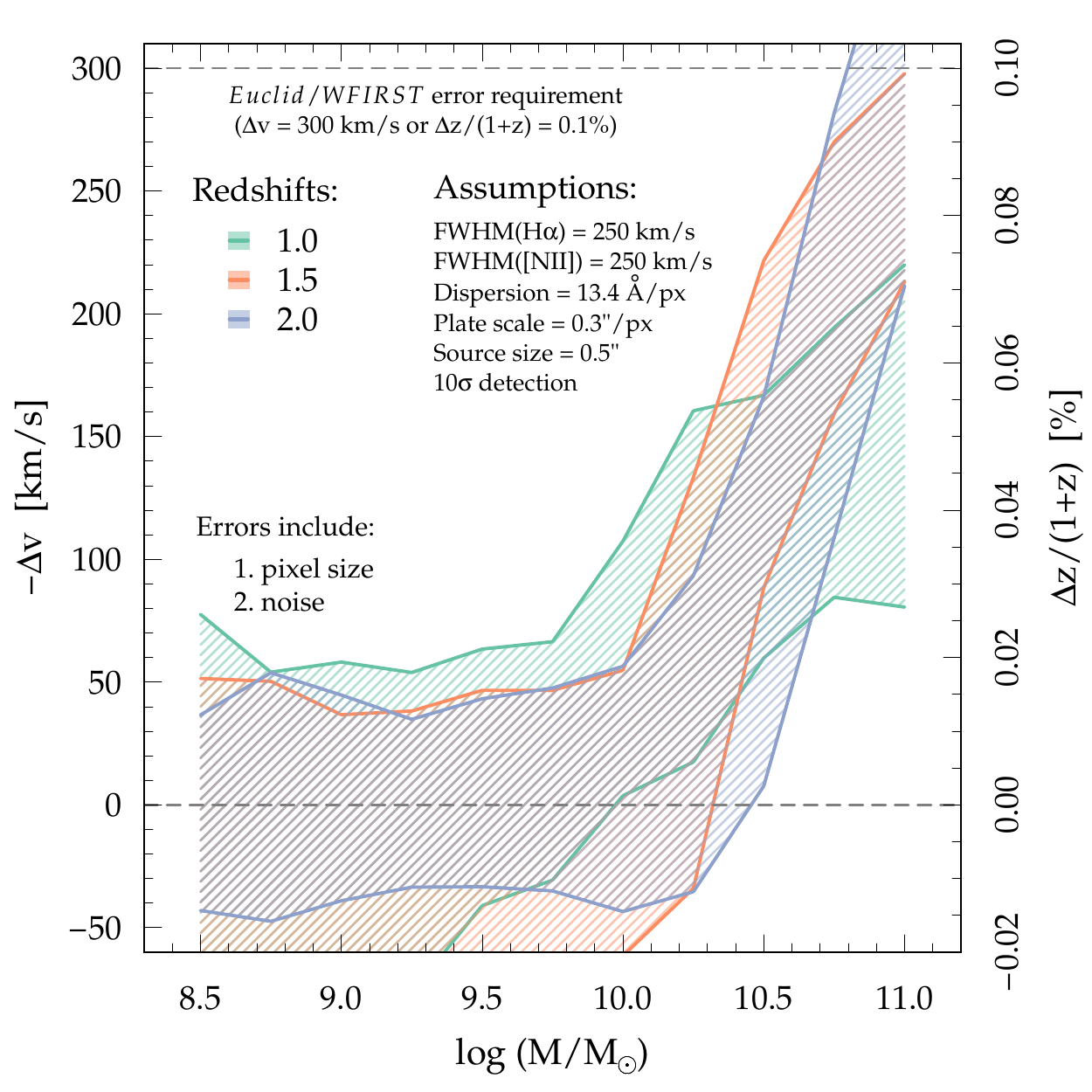}\\
\caption{Velocity shifts and redshift biases in the determination of the \halpha\ wavelength centroid due to blending with \nii\ for a source size of $0.5\arcsec$ and S/N$=10$ on the line. Shown are simulations for a range of stellar masses at $z=1$ (green), $1.5$ (orange), and $2.0$ (blue). The \halpha\ centroid is measured by a Gaussian fit to the observed (i.e., resolution adjusted) spectrum at $6500-6600\,{\rm \AA}$ in rest-frame (corresponding to $7-11$ \emph{Euclid} pixel-pairs at $1 < z < 2$). The hatched area shows the uncertainty due to \emph{Euclid}'s finite spectral resolution and measurement noise.
	Our simple simulation suggests that the velocity shifts are less than the error requirement for \emph{Euclid} and \emph{WFIRST} ($\Delta v = 300\,{\rm km\,s}^{-1}$ or $\Delta z / (1+z) = 0.1\%$). The bias is increasing steeply with stellar mass at $\logm \gtrsim10$ due to an increasing \niiha\ ratio.
\label{fig:centroid}}
\end{figure}

	\subsection{Spectroscopic redshift measurements from blended \halpha\ and \nii\ } \label{sec:centroid}
    
	The blending of the \halpha\ and \nii\ lines can result in biases in the determination of spectroscopic redshifts.
    Here, we study this bias as a function of stellar mass and redshift via a simple preliminary simulation based on the predictions from our model. 
    
    We approximate each of the three emission lines by a Gaussian and assign fluxes relative to the \halpha, namely, F(\nii$\lambda6584) = $\niiha$\times {\rm F(H\alpha)}$, and F(\nii$\lambda6548) = \frac{1}{3}\,{\rm F}($\nii$\lambda6584$). The flux ratio \niiha\ is computed from our Equation~\ref{eq:final} and we use vacuum wavelengths for the emission lines ($6549.86\,{\rm \AA}$, $6564.61\,{\rm \AA}$, and $6585.27\,{\rm \AA}$ for \nii$\lambda6548$, \halpha, and \nii$\lambda6584$, respectively). For the full width at half maximum (FWHM) of the lines we assume $250\,{\rm km\,s}^{-1}$ as commonly measured on average by spectroscopic surveys at $\logm = 10$. Our final results do not significantly depend on the exact values for the FWHMs, mainly because of the somewhat low spectral resolution of \emph{Euclid}.
    
    We then convolve and bin this input spectrum to the \emph{Euclid} resolution and pixel size, assuming a spectral dispersion of $13.4\,{\rm \AA/px}$ and a plate scale of $0.3\,\arcsec/{\rm px}$ \citep{VAVREK16}. The spectral dispersion results in an $R=\lambda / \Delta\lambda$ of $490-735$ for \halpha\ at $1 < z < 2$ in a $2-$pixel resolution element for a source of $0.3\arcsec$ diameter.
    In the following, we assume a more realistic source diameter of $0.5\arcsec$ and a point spread function (PSF) FWHM of $0.4\arcsec$ \citep{VAVREK16}, which decreases the resolution by a factor $\sim2$, and bin the final observed spectrum to a $2-$pixel resolution element. Note that \emph{Euclid}'s resolution element decreases proportional to $(1+z)$ for increasing redshift, i.e., $13.4/(1+z)\,{\rm \AA/px}$, as the observed spectrum stretches in wavelength. This results in an increase in resolution of $33\%$ from $z=1$ to $z=2$ for a galaxy of fixed apparent size.
    We also add noise to the output spectrum according to the required $3.5\sigma$ flux limit of $2\times 10^{-16}\,\ergPerSecondPerCM$, which we assume is the integrated flux over the blended \halpha\ and \nii\ emission lines. In the following, we assume a source detected at $10\sigma$ in integrated line flux. This is a good approximation for most of \emph{Euclid}'s detected sources according to predictions of the \halpha\ luminosity function (see Section~\ref{sec:half}).
    
    Our preliminary results indicate that \emph{Euclid} will generally not resolve \halpha\ and \nii\ for most of the assumed \niiha\ values (and thus stellar masses) for a source of $0.5\arcsec$ and a S/N of 10, however, an asymmetry of the blended line caused by the \nii\ red-ward of \halpha\ is identifiable.
We therefore compute the \halpha\ centroid on the final convolved and binned spectrum by fitting a Gaussian at $6500-6600\,{\rm \AA}$ in rest-frame. This wavelength width encompasses both \nii\ as well as \halpha\ and corresponds to $7-11$ \emph{Euclid} pixel-pairs ($26.8{\rm \AA}$ per pair) at $1 < z < 2$.

    Figure~\ref{fig:centroid} shows the resulting centroid shifts in velocity and redshift with respect to the true \halpha\ wavelength as a function of stellar mass at $z=1.0$, $1.5$, and $2.0$. The hatched area combines the errors from the noise and the finite pixel size of \emph{Euclid}. The latter is obtained by shifting the binning of the final spectrum by up to half a resolution element.
    The redshift bias increases towards higher stellar masses due to the larger \niiha\ flux ratio\footnote{The \niiha\ flux ratio is $0.05$ ($0.03$) and $0.40$ ($0.63$) in linear scaling for a galaxy with $\logm = 8.5$ and $\logm = 11.0$, respectively, for $z=1$ ($z=2$).}. Furthermore, the bias decreases slightly with increasing redshift at a fixed stellar mass due to the increasing resolution with increasing redshift. While this effect is only small for a source of S/N$=10$, we would expect a much larger reduction of the bias at higher S/N, where \nii\ and \halpha\ will likely be resolved at the highest redshifts. However, the amount of detected sources at S/N$>10$ and high redshift is likely small (see Section~\ref{sec:half}).

     In general, we find velocity shifts that are better than the error requirement for \emph{Euclid}, which is $\Delta v = 300\,{\rm km\,s}^{-1}$ or $\Delta z / (1+z) = 0.1\%$ \citep{VAVREK16}.
    For a galaxies at $\logm < 10$, we expect negligible biases ($|\Delta v| < 50\,{\rm km\,s}^{-1}$), however, the biases increase sharply at $\logm > 10$ due to the increasing \niiha\ ratio.
    For a galaxy of $\logm = 11$, we expect significant biases around $100-300\,{\rm km\,s}^{-1}$, or $\Delta z / (1+z) \sim 0.04-0.10\%$.
     If uncorrected, such a shift will introduce a bias in the BAO measurements in the radial direction. Specifically, at $z=1$ ($z=1.5$), a shift of $\Delta z / (1+z)=0.04-0.10\%$ corresponds to $1.0-2.4\,{\rm Mpc}$ ($0.7-1.8\,{\rm Mpc}$) or roughly $0.6-1.6\%$ ($0.5-1.2\%$) of the BAO scale at $\sim 150\,{\rm Mpc}$.
     This is significant since the BAO peak itself is a few-percent level signal in the galaxy correlation function that needs to be measured at the precision of a few percent or better.
    Finally, we note that the \niiha\ blending may lead to additional systematic effects for BAO/RSD measurements, if metallicity evolution is correlated with density. This will be examined further in future studies.

   Our preliminary simulation is very basic and we will use more realistic grism simulations in the future for more detailed investigations. Furthermore, the evolution and distribution of the angular sizes of the galaxies should be taken into account (the combined effect of the increasing cosmological angular diameter distance and the decreasing physical size of the galaxies with redshift).

    Finally, we note that \halpha\ and \nii\ emission lines will be likely resolved for many galaxies detected by \emph{WFIRST} at its spectral resolution of $R\sim600-900$ for \halpha\ at $1 < z < 2$ \citep[with a dispersion of $10.85\,{\rm \AA/px}$,][]{SPERGEL15}, therefore much smaller biases are expected.

\begin{figure*}
\centering
\includegraphics[width=2.1\columnwidth, angle=0]{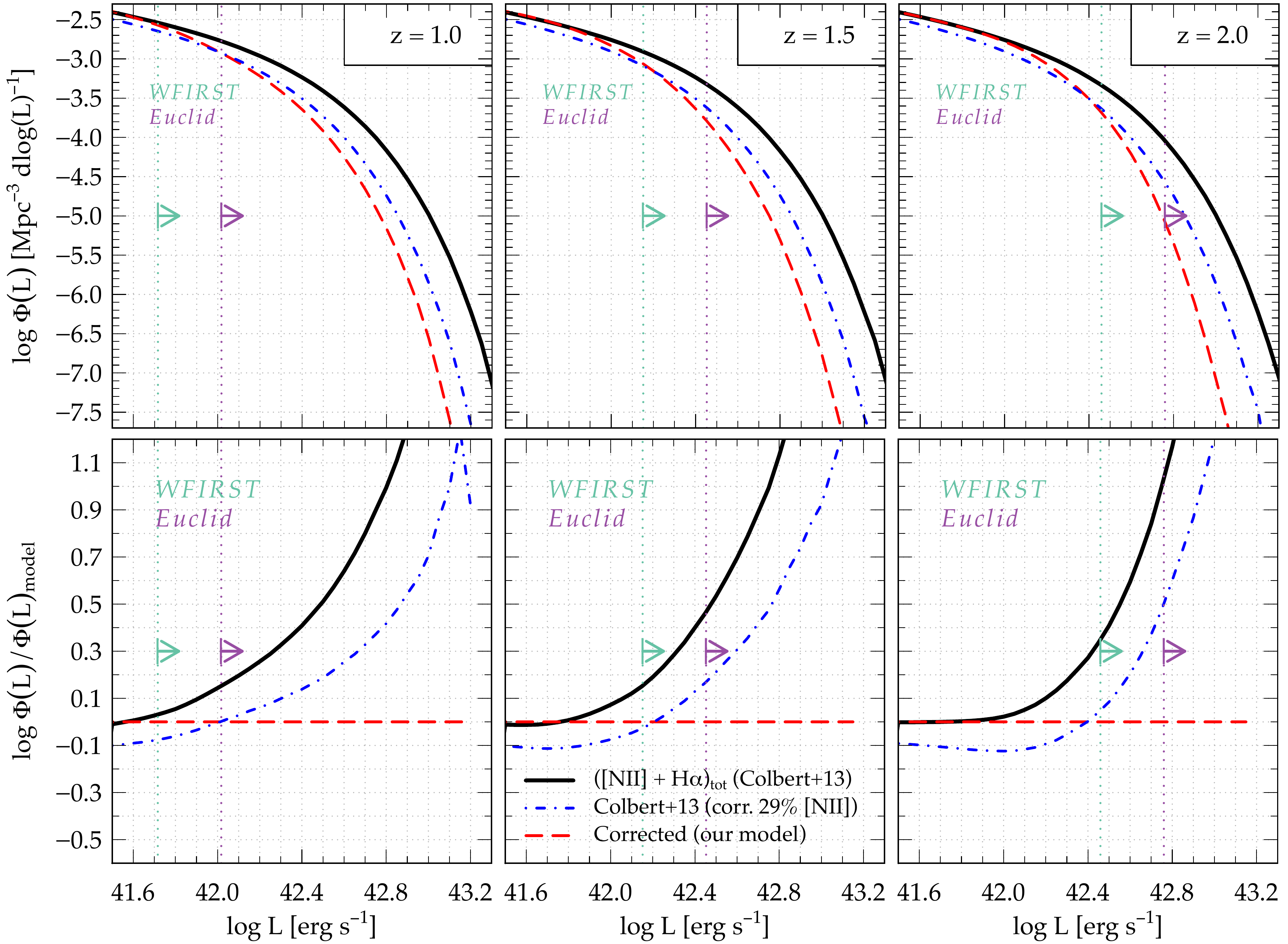}\\
\caption{Differential effects of redshift- and stellar mass-dependent total \nii\ flux contamination corrections on the number counts of \halpha\ emitters displayed on the example of the \citet{COLBERT13} luminosity function at $0.9 < z < 1.5$ ($\langle z \rangle = 1.2$).
\textit{Top panels:} The black solid line shows the total \niiphatot\ LF observed by \citet{COLBERT13} at $0.9 < z < 1.5$ ($\Phi_{\rm obs}$) and the blue dot-dashed line is the \halpha\ LF derived from a correction assuming a constant total \nii\ flux contamination fraction of $29\%$ ($\Phi_{\rm 29}$) by the same authors. The red dashed line shows the \halpha\ LF derived from the \niiphatot\ LF using our model for \nii\ contamination ($\Phi_{\rm model}$). The luminosity limits for \emph{WFIRST} and \emph{Euclid} are shown as green and purple arrows, respectively.
\textit{Bottom panels:} The three LFs relative to our \nii\ corrected \halpha\ LF $\Phi_{\rm model}$ with the same color-code as in the top panels.
A redshift and stellar mass dependent total \nii\ flux contamination fraction is important to obtain accurate \halpha\ emitter number counts.
\label{fig:half}}
\end{figure*}
 
    \subsection{Impact on \halpha\ luminosity function and number count predictions for \emph{Euclid} and \emph{WFIRST}}\label{sec:half}
    
    The dark energy figure-of-merit for both \emph{WFIRST} and \emph{Euclid} is very sensitive to the number density of \halpha\ emitting galaxies.
    
    Measurements of the observed \emph{blended} \halpha\ LF of low-resolution \emph{HST} grism surveys are used to predict the observed number counts for future large surveys such as \emph{WFIRST} or \emph{Euclid} \citep{COLBERT13,MEHTA15}.
    While these number counts are accurate for the redshift, stellar mass, and sSFR distribution of the grism surveys, any extrapolation beyond that to match \emph{WFIRST}'s and \emph{Euclid}'s  parameter space requires the knowledge of the intrinsic \halpha\ LF and therefore an accurate assessment of the total \nii\ flux contamination fraction \citep[see also discussion in][]{POZZETTI16}.
    
    Importantly, future large area galaxy surveys will be predominantly probing the bright-end of the LF at its exponential decline and therefore any uncertainty in the brightest \halpha\ luminosities will have a significant impact on the \halpha\ number counts. Furthermore, the derivation of the intrinsic \halpha\ LF will be important for studying many physical properties of the galaxies, such as their SFRs. This does not only apply to future studies but also current grism spectroscopy and narrow-band photometric observations that do not resolve \nii\ and \halpha.
    Here we investigate the \textit{relative} change in the intrinsic \halpha\ emitter number counts when using different corrections for \nii\ contamination. Specifically, we study \textit{(i)} a constant $29\%$ contamination and \textit{(ii)} the stellar mass and redshift dependent \niiha\ flux ratios predicted by our model (Equation~\ref{eq:final}).
    
	To derive the intrinsic \halpha\ LF using our model \nii\ contamination ($\Phi_{\rm model}$), we start with the \textit{observed} \niiphatot\ LF ($\Phi_{\rm obs}(L)$) measured by \citet{COLBERT13} at $0.9 < z < 1.5$. We obtain this LF from their published \halpha\ LF ($\log \Phi_* = -2.70$, $\log L_* = 42.18$, and $\alpha = -1.43$) by dividing the luminosities by a factor of $(1-0.29)$ to undo \textit{constant} total \nii\ flux contamination fraction correction of $29\%$, which the authors applied.
	In the following, we treat $\Phi_{\rm obs}$ as the true observed LF. Importantly, this LF is not corrected for \nii\ contamination and dust. $\Phi_{\rm obs}$ is redshift dependent because of the evolution of the star forming main-sequence, but here we do not model this dependence across $z=1-2$ as we are only interested in the effects of \nii\ contamination and not the absolute number of galaxies. On the other hand, the stellar mass and redshift dependent \nii\ contamination correction will change the \textit{intrinsic} \halpha\ LF across the redshift range studied here.
	To obtain $\Phi_{\rm model}$, we choose an approach, which only uses the measured \niiphatot\ luminosities as input and assumes the most likely underlying stellar mass distribution (robustly determined from other studies) from which we obtain the \nii\ contamination from our model. This approach has the advantage that it enables an easy implementation and propagation of a variety of uncertainties into final results. Furthermore, this method results in reliable intrinsic \halpha\ LFs even if the mass distribution is poorly measured due to the lack of sufficient multi-wavelength data, as long as the selection function of the galaxy sample is known.
	Here, we make use of the \citet{SCHREIBER15} parameterization of the star-forming main-sequence to derive the underlying stellar mass distribution (we comment below on possible shortcomings). We note that the choice of different parameterizations \citep[e.g.,][]{TOMCZAK16} should not change the following results.
	We start with a distribution of \niiphatot, which we sample from $\Phi_{\rm obs}$. To obtain stellar masses for these galaxies, we use a ``backwards engineering'' technique. First, we convert the SFRs of the \citet{SCHREIBER15} parameterization into \halpha\ luminosities using the \citet{KENNICUTT98} prescription. Thereby we include a dispersion of $0.3\,{\rm dex}$ measured on the SFR vs. stellar mass main-sequence. Second, we redden the \halpha\ luminosities according to the relation between $A_{\rm H\alpha}$ (\halpha\ extinction) and stellar mass robustly derived from the spectra of local galaxy samples in SDSS \citep{GARN10}. This relation holds for the WISPS sample at $0.8 < z < 1.5$ as shown in \citet{DOMINGUEZ13}. Third, we add the contribution of \nii\ to the \halpha\ luminosity by using our model. Finally, this translation between dust reddened \niiphatot\ luminosities and stellar masses allows us to obtain the underlying stellar mass distribution and intrinsic \halpha\ luminosities for the galaxy sample describing $\Phi_{\rm obs}$, from which we are now able to re-compute the intrinsic \halpha\ LF $\Phi_{\rm model}$.

    Figure~\ref{fig:half} shows the three LFs; \textit{(i)} the total \niiphatot\ LF ($\Phi_{\rm obs}$), \textit{(ii)} the \halpha\ LF corrected with constant $29\%$ \nii\ correction as published in \citet{COLBERT13} ($\Phi_{\rm 29}$), and \textit{(iii)} the \halpha\ LF with redshift and stellar mass dependent \nii\ correction from our model ($\Phi_{\rm model}$). The LFs are shown in absolute values (top panels) and relative to $\Phi_{\rm model}$ (bottom panels) at redshifts $z=1$, $z=1.5$, and $z=2$. The luminosity limits for \emph{WFIRST} and \emph{Euclid} (redshift dependent) are shown as arrows for reference.
    Note that only $\Phi_{\rm model}$ changes with redshift due to the redshift-dependent \niiha\ flux ratio, while the other LFs are unchanged.
       
    First of all, it is evident that an accurate \nii\ correction is crucial at the bright-end of the LF where the number counts exponentially drop and the LF is dominated by massive galaxies with large \nii\ corrections (see Figure~\ref{fig:n2massz}).
    Similarly, the difference between $\Phi_{\rm 29}$ and $\Phi_{\rm model}$ increases towards the bright-end of the LF due to its steepness and the mass dependence of the \nii\ correction. While at $\log(L_{\rm H\alpha}) < 42.4$ a constant \nii\ contamination correction generally underestimates the \halpha\ number counts by $\lesssim 0.1\,{\rm dex}$ with respect to our model, at higher luminosities the deviation is more severe.
    For example, at $z=1.5$ the number counts of galaxies at $\log(L_{\rm H\alpha}) = 43.0$ ($\logm \sim 10.6$) would be overestimated by approximately a factor of $8$ ($0.9\,{\rm dex}$) with respect to using a mass and redshift dependent \nii\ contamination. This factor is expected to be less (factor $5$, $0.7\,{\rm dex}$) at $z=1$ and more (factor $15$, $1.2\,{\rm dex}$) at $z=2$ at the same \halpha\ luminosity.
    Such biases are not to be neglected as \emph{Euclid} will probe the high-luminosity part of the \halpha\ LF as indicated by the purple arrows in Figure~\ref{fig:half}.     
    
    Our Equation~\ref{eq:final} was also used in \citet{MERSON18} to transform the \nii\ blended \halpha\ flux in the WISPS data into true \halpha\ fluxes for calibrating the semi-analytical galaxy formation code \textsc{Galacticus} \citep{BENSON12}, so that reliable forecasts of galaxy number counts can be obtained for the galaxy redshift surveys planned for \emph{Euclid} and \emph{WFIRST}.
    
    To conclude, we briefly discuss possible caveats of our approach.
    First, we note that the emission line selected WISPS galaxies may probe a different stellar mass distribution as in the \citet{SCHREIBER15} study (which is based on $H$ and $K$ band continuum selected galaxies). Specifically, we would expect the average stellar mass at a given SFR to be lower in the case of emission line selected galaxies \citep[e.g., ][]{LY12}, hence our stellar masses would be overestimated. Assuming conservatively a factor of two lower average stellar masses per SFR would lead to $<30\%$ lower \nii\ contamination over the mass range $9.5 < \logm < 11.0$ (approximately $41.9 < \log(L_{\rm H\alpha}) < 43.5$) at $z=1.5$. This translates into $<0.15\,{\rm dex}$ less overestimation of the \halpha\ emitter counts if using no or a constant $29\%$ \nii\ contamination correction compare to our model. This is negligible compared to the large corrections needed at the bright end of the LF.
    Second, we note that the \citet{KENNICUTT98} relation to obtain SFRs from \halpha\ luminosities was derived from galaxies with solar metallicity and an electron temperature of the ionized gas of $10^4\,{\rm K}$. These assumptions may not be valid at high redshifts. Using the metallicity dependent parameterization of the Kennicutt relation by \citet{LY16b}, we estimate that the SFR for a given \halpha\ luminosity is $\sim0.2\,{\rm dex}$ lower for galaxies at 1/5$^{\rm th}$ of solar metallicity. As above, this would lead to similar or less overestimation of stellar mass and \nii\ contamination, respectively, and therefore mostly negligible modifications to our results.

    Finally, it should be mentioned that the unknown contribution of AGNs at high stellar masses and redshift (see also Section~\ref{sec:agns}) adds an additional uncertainty to the \halpha\ LF that can have similar impacts as inadequate \nii\ contamination corrections. Specifically, \citet{GENZEL14} find that two-thirds of their sample of $z \sim 1-3$ galaxies above $\logm = 10.9$ shows broad nuclear emission that could potentially be explained by the occurrence of an AGN. In this case, this would add almost a factor of three ($\sim 0.5\,{\rm dex}$) uncertainty on the number counts at $\logm > 10.9$ (approximately $\log(L_{\rm H\alpha}) > 43.0$ at $z=1.5$).
    Hence, this will clearly dominate the uncertainties of the \halpha\ LF at high stellar masses (in comparison, the uncertainties from our model add up to about $10-20\%$). However, compared to the difference in the \halpha\ LF between a constant and our model based \nii\ correction, the uncertainty due to AGN contamination is a factor of two lower ($0.5\,{\rm dex}$ compared to $\sim0.9\,{\rm dex}$ at $\log(L_{\rm H\alpha}) = 43.0$ at $z=1.5$, see Figure~\ref{fig:half}).

\vfill\null 
\section{Summary and Outlook}\label{sec:end}

\subsection{Summary}

	We present a parameterization of the \niiha\ flux ratio as a function of stellar mass and redshift from $0 < z < 2.7$ for stellar masses of $8.5 < \logm \lesssim 11.0$.
	Our model encompasses the shift in the BPT locus defined by observed high-redshift data and the dependence of stellar mass on the BPT diagram on the \niiha\ and \oiiihb\ emission line ratios from local galaxies.
    Our description is easily applicable to simulations for modeling \nii\ emission, current low-resolution grism and narrow-band observations to derive intrinsic \halpha\ fluxes, and to forecast the \halpha\ emission line galaxy number counts of future surveys.
    
    We find large variations in the total \nii\ flux contamination fraction at a fixed redshift due to its dependency on stellar mass. Hence, we emphasize three main implications on current data as well as future surveys.
    
	\begin{itemize}
	\item The use of a constant \nii\ flux contamination fraction over- and under-predicts the true \nii\ contamination mainly as a function of stellar mass and redshift. This can lead to severe mass and redshift dependent biases in the determination of the intrinsic \halpha\ LF as wells other physical parameters computed from it. For example, a constant \nii\ contamination of $29\%$ overestimates the true value for galaxies at $\logm \lesssim 10$ at $z>0.5$ by a factor of up to $3$.
    
     \item Intrinsic \halpha\ emitter number counts based on current \emph{HST} grism surveys assuming a constant \nii\ flux contamination fraction of $29\%$ are likely overestimated by $0.9\,{\rm dex}$ (factors of $8$) and more at observed $\logL > 43.0$ at $z=1.5$. Hence, the extrapolation of the \emph{observed} \niipha\ number counts from these studies to match future surveys such as \emph{WFIRST} and \emph{Euclid}, which probe different redshift and stellar mass distributions, requires a redshift and stellar mass dependent modeling of the \nii\ flux contamination fraction as presented here.
     
    \item The blending of \halpha\ and \nii\ leads to a mass- and redshift-dependent systematic bias in the redshift measurement for \emph{Euclid}. Our preliminary simulations indicate a redshift bias $\Delta z/(1+z) \sim 0.04-0.10\%$ for the most massive galaxies. This leads to a systematic bias of $0.5-1.6\%$, depending on redshift, in the BAO scale measurement in the radial direction at $150\,{\rm Mpc}$.

	\end{itemize}

\subsection{Outlook}

	To examine our results in the context of current galaxy formation theory we plan to compare our model predictions to the predictions from a semi-analytical galaxy formation model (for example the \textsc{Galacticus} model). Such a comparison would allow us to further investigate the dependence of the \niiha\ ratio on additional intrinsic galaxy properties, including the sSFR, as well as help test the validity of our model for redshifts $z\gtrsim3$.

	We will deepen our study on the spectroscopic redshift measurement biases by using realistic grism simulations, by applying more accurate noise levels expected for \emph{Euclid}, and by including statistically more detailed properties of the galaxies (such as varying physical size). In addition, the application of our model to large area mock catalogs would allow further examination of how a redshift bias will impact determination of the BAO peak position, as well as subsequent cosmological parameter estimation. In particular, we will study techniques for correcting this redshift bias.
    
    Finally, we stress that further observational follow-up is needed to tighten our model especially at the massive end. Specifically, a \emph{WFC3} grism filler program targeting massive ($\logm > 11$), star forming galaxies at $z\gtrsim2$ would be useful to understand the line ratios of massive galaxies as well as the contribution of broad-line emission and AGN.

\acknowledgements
The authors thank J. Colbert for helpful comments and discussions. We also thank the referee for the very useful comments, which helped to improve this paper. A.M. acknowledges sponsorship of a NASA Postdoctoral Program Fellowship. A.M. was supported by JPL, which is run under contract by California Institute of Technology for NASA.

\appendix

\section{SQL commands for SDSS galaxy selection}\label{app:sql}
In the following, we list the SQL commands that were used to retrieve our SDSS sample from \url{http://skyserver.sdss.org/dr12/en/tools/search/sql.aspx}.
\vspace{0.5cm}\\
\texttt{
SELECT top 200000\\
\indent p.ObjID, s.fiberID,\\\
\indent p.ra, p.dec, s.z,\\
\indent p.modelMag\_u, p.modelMag\_g,\\
\indent p.modelMag\_r, p.modelMag\_i, \\
\indent p.modelMag\_z,\\
\indent p.expRad\_r, p.expRad\_i,\\
\indent e.sfr\_tot\_p50,\\
\indent e.lgm\_tot\_p50,\\
\indent g.h\_alpha\_eqw,\\
\indent g.oii\_3726\_flux,\\
\indent g.oii\_3726\_flux\_err,\\
\indent g.oii\_3729\_flux,\\
\indent g.oii\_3729\_flux\_err,\\
\indent g.neiii\_3869\_flux,\\
\indent g.neiii\_3869\_flux\_err,\\ 
\indent g.h\_beta\_flux,\\
\indent g.h\_beta\_flux\_err,\\
\indent g.oiii\_5007\_flux,\\
\indent g.oiii\_5007\_flux\_err,\\
\indent g.oi\_6300\_flux, \\
\indent g.oi\_6300\_flux\_err,\\
\indent g.h\_alpha\_flux,\\
\indent g.h\_alpha\_flux\_err,\\
\indent g.nii\_6584\_flux,\\
\indent g.nii\_6584\_flux\_err, \\
\indent g.sii\_6717\_flux,\\
\indent g.sii\_6717\_flux\_err, \\
\indent g.sii\_6731\_flux,\\
\indent g.sii\_6731\_flux\_err \\
FROM photoObj p\\
JOIN specObj s ON s.bestObjID = p.objID\\
JOIN galSpecLine g ON g.specObjID = s.specObjID    \\
JOIN galSpecExtra e ON e.specObjID = g.specObjID     \\
WHERE\\
\indent s.class = 'galaxy' \\
\indent and e.bptclass = 1\\
\indent and s.zWarning = 0 \\
\indent and g.h\_alpha\_flux / nullif(g.h\_alpha\_flux\_err,0) $>$ 5\\
}

\section{Scatter in \niiha\ ratios at $z\sim1.6$ and $z\sim0$}\label{app:scatter}

\begin{figure*}
\centering
\includegraphics[width=0.49\columnwidth, angle=0]{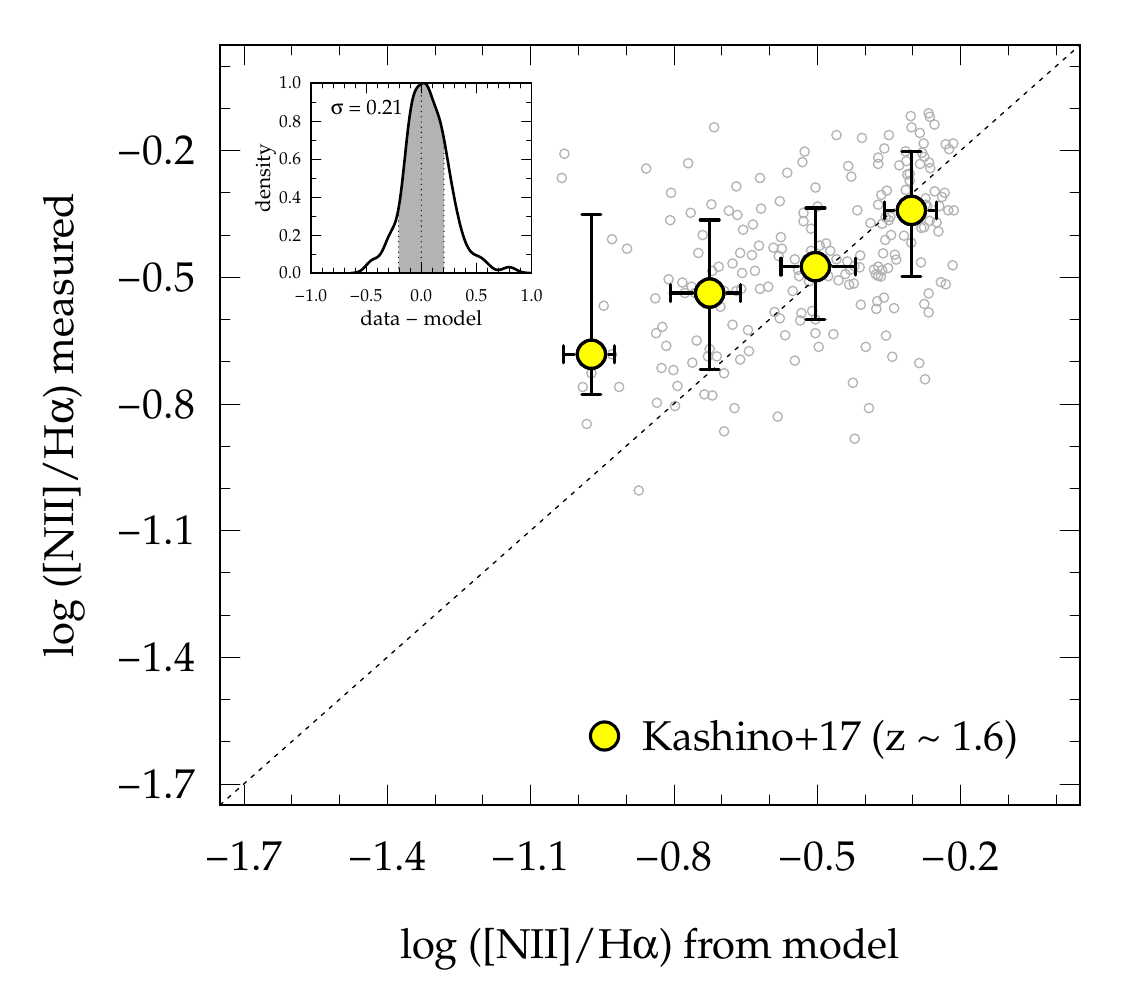}
\includegraphics[width=0.49\columnwidth, angle=0]{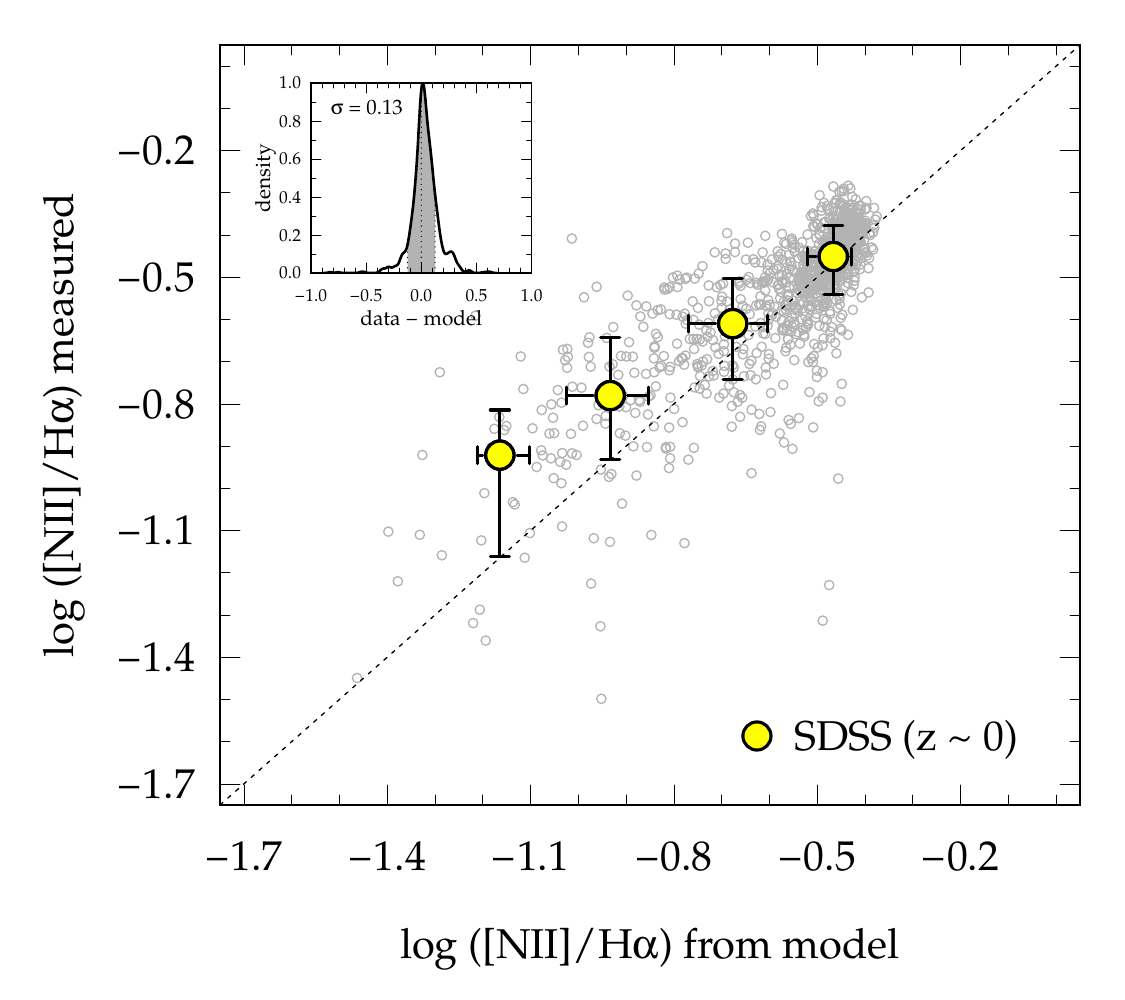}
\caption{Same as Figure~\ref{fig:n2scatter} but for $z\sim1.6$ \citep[left, ][]{KASHINO17} and SDSS galaxies (right). 
\label{fig:n2scatter2}}
\end{figure*}

Figure~\ref{fig:n2scatter} compares the true (i.e., measured) \niiha\ ratios to the ones provided by our model at $z\sim2.3$. In Figure~\ref{fig:n2scatter2}, we show the same figure for $z\sim1.6$ and $z\sim0$ for reference. The scatter between true and model \niiha\ ratios is $0.21\,{\rm dex}$ for $z\sim 1.6$ (similar to $z\sim2.3$) and $0.13\,{\rm dex}$ for $z\sim0$.

\bibliographystyle{aasjournal}
\bibliography{bibli.bib}





\end{document}